\documentclass[%
 reprint,
superscriptaddress,
 amsmath,amssymb,
 aps,
pra,
]{revtex4-1}

\usepackage{amsthm,amssymb}
\usepackage{amsmath}
\usepackage{dcolumn}% Align table columns on decimal point
\usepackage{bm}% bold math
\usepackage{multirow} % multirow
\usepackage{mathdots}
\usepackage{enumerate}
\usepackage{algorithm}
\usepackage{algorithmicx}
\usepackage{subfig}
\usepackage{graphicx}    % 用于插入图片

\usepackage{algpseudocode}
\usepackage{textcomp}
\usepackage{color}

\usepackage{epstopdf}
\usepackage{epsfig}
\usepackage{array}
\usepackage{booktabs}
\usepackage{setspace}%
\usepackage{amsmath}%
\usepackage{tikz}%
\usepackage{hyperref}
\usepackage{multirow}
\usepackage{changes}
\usepackage{booktabs}
\usepackage{tabularx}
\usepackage{xcolor}

\usetikzlibrary{shapes,arrows}
\usepackage{caption} % 确保加载 caption 宏包

\usepackage{hyperref}
\hypersetup{
    colorlinks=true,
    linkcolor=blue,      % 设置链接颜色
    citecolor=blue,      % 设置引用文献颜色
    urlcolor=blue        % 设置超链接颜色
}

\theoremstyle{remark}

\begin{document}

\makeatletter
\newcommand{\ud}{\mathrm{d}}
\newcommand{\rmnum}[1]{\romannumeral #1}
\newcommand{\polylog}{\mathrm{polylog}}
\newcommand{\ket}[1]{|{#1}\rangle}
\newcommand{\bra}[1]{\langle{#1}|}
\newcommand{\inn}[2]{\langle{#1}|#2\rangle}
\newcommand{\Rmnum}[1]{\expandafter\@slowromancap\romannumeral #1@}
\newcommand{\udots}{\mathinner{\mskip1mu\raise1pt\vbox{\kern7pt\hbox{.}}
        \mskip2mu\raise4pt\hbox{.}\mskip2mu\raise7pt\hbox{.}\mskip1mu}}
\makeatother

\preprint{APS/123-QED}

\title{An Adaptive Mixer Allocation Strategy for the Quantum Alternating Operator Ansatz}

\author{Xiao-Hui Ni}
\affiliation{State Key Laboratory of Networking and Switching Technology, Beijing University of Posts and Telecommunications, Beijing 100876, China}
\affiliation{School of Cyberspace Security, Beijing University of Posts and Telecommunications, Beijing, 100876, China}
\author{Yu-Sen Wu}
\affiliation{School of Artificial Intelligence, Beijing Normal University, Beijing 100875, China}
\author{Bin-Bin Cai}
\affiliation{College of Computer and Cyber Security, Fujian Normal University, FuZhou ,350117, China}
\author{Wen-Min Li}
\affiliation{State Key Laboratory of Networking and Switching Technology, Beijing University of Posts and Telecommunications, Beijing 100876, China}
\author{Su-Juan Qin}
\email{qsujuan@bupt.edu.cn}
\affiliation{State Key Laboratory of Networking and Switching Technology, Beijing University of Posts and Telecommunications, Beijing 100876, China}
\author{Fei Gao}
%\email{gaof@bupt.edu.cn}
\affiliation{State Key Laboratory of Networking and Switching Technology, Beijing University of Posts and Telecommunications, Beijing 100876, China}

\date{\today}

\begin{abstract}
	Recently, Hadfield et al. proposed the quantum alternating operator ansatz algorithm (QAOA+), an extension of the quantum approximate optimization algorithm (QAOA), to solve constrained combinatorial optimization problems (CCOPs). Compared with QAOA, QAOA+ enables the search for optimal solutions within a feasible solution space by encoding problem constraints into the mixer Hamiltonian, thereby reducing the search space and eliminating the possibility of yielding infeasible solutions. However, QAOA+ may incur high overall gate costs when the mixer is applied to all qubits in each layer, and each mixer is costly to implement. To address this challenge, an adaptive mixer allocation strategy is tailored for QAOA+. The resulting algorithm, which integrates this strategy into the original QAOA+ framework, is referred to as AMA-QAOA+. Unlike QAOA+, AMA-QAOA+ adaptively applies the mixer to a subset of qubits in each layer of the mixer unitary operator based on an evaluation function. The performance of AMA-QAOA+ is evaluated on the maximum independent set problem. Numerical simulation results show that, under the same number of optimization runs, AMA-QAOA+ achieves better solution quality than QAOA+, with the optimal approximation ratio improved by $5.30\%$ on Erdős–Rényi random graphs and $5.41\%$ on 3-regular graphs. Moreover, AMA-QAOA+ significantly reduces the CNOT gate consumption, requiring only $15.30\%$ and $25.18\%$ of the CNOT gates used by QAOA+ on Erdős–Rényi and 3-regular random graphs, respectively. These results demonstrate that AMA-QAOA+ enhances solution quality and computational efficiency, enabling the design of more compact and resource-efficient quantum circuits.
	
\end{abstract}

\pacs{Valid PACS appear here}
\maketitle

\section{Introduction}

% QAOA应用广泛、局限性
The quantum approximate optimization algorithm (QAOA) \cite{QAOA, RQAOA, XY_QAOA} is a representative quantum-classical hybrid algorithm and has been widely applied to solve various combinatorial optimization problems (COPs) \cite{INTERP,angle_conjecture, MLI, TQA, QIRO, IQO, exact_cover, exact_cover1, minimum_vertex_cover}. When solving the constrained COPs (CCOPs), QAOA tends to encode the problem constraints into the target Hamiltonian utilizing the penalty terms \cite{QIRO, IQO, minimum_vertex_cover, Ising, TSPTW}, and then search for the quasi-optimal solution using a parameterized quantum circuit (PQC). However, this approach may lead to infeasible solutions (i.e., solutions that violate one or more problem constraints), as QAOA explores the entire Hilbert space, which includes both feasible and infeasible solutions \cite{RUAN}.

\medskip
%关键点： 1.QAOA+的原理， 2.QAOA+应用广泛 3. QAOA+存在的缺陷以及原因，造成的后果。  
To address this issue, Hadfield et al. \cite{QAOA+} proposed a novel quantum alternating operator ansatz algorithm (QAOA+). QAOA+ constructs a feasible solution space by encoding the problem constraints into the mixer Hamiltonian and searches for the quasi-optimal solution within this space \cite{RUAN}. This approach reduces the search space and eliminates the possibility of yielding infeasible solutions \cite{minimum_exact_cover}. Leveraging these advantages, QAOA+ has been widely utilized to solve various problems \cite{minimum_exact_cover, CVRP, QLS, PQA, MIS&graph_coloring, UFLP,protein_folding}. However, QAOA+ may require a large number of quantum gates, especially when the mixer is applied to all qubits in implementing each layer of the mixer unitary operator and the implementation gate overhead of each mixer is costly \cite{QLS,gate_decomposition_ieee,gate_decomposition_quantum}. Such high resource demands limit the size and complexity of problems that can be addressed on current \textcolor{black}{resource-constrained quantum} devices. Therefore, it is necessary to search for algorithms to reduce the expensive gate overheads brought by \textcolor{black}{full-mixer allocation strategy} for QAOA+.

% 第四段：现有的可以用来降低混合器数量的方法。方法原理，以及存在的问题。
% 在现有研究中，已有一些为变分量子特征求解器（VQE）和QAOA量身定制的自适应算法，旨在构造更加紧凑的Ansatz线路。这些自适应算法的核心思想是迭代地扩展Ansatz线路。具体来说，算法从预先定义好的算子池中选择一个使得评估函数最大的量子门，并将其添加到线路中，然后对整体线路进行参数优化。算法通过重复线路扩展和参数优化的过程，直至目标函数收敛。然而，现有自适应算法存在一些局限性。首先，它们的评估方式通常较为单一，主要依赖于某一组参数的梯度进行评估，可能无法准确或全面地捕捉到优化所需要的不同信息。其次，算法需要多轮优化，这会增加迭代次数，从而影响其效率和适用性。
\medskip
In existing research, several adaptive strategies designed for variational quantum eigensolver (VQE) \cite{adaptive_VQE, qubit_adaptive_vqe} and QAOA \cite{Adaptive_qaoa, dynamic_adaptive_qaoa} have been proposed to construct more compact ansatz circuits by reducing redundant gates, where the VQE algorithm using an adaptive strategy is called the Adaptive-VQE algorithm, and the QAOA algorithm with an adaptive strategy is referred to as the Adaptive-QAOA algorithm. These adaptive algorithms progressively expand the ansatz circuit starting from a shallow initial circuit. At each expansion, \textcolor{black}{an operator} that maximizes the evaluation function is initially selected from a predefined operator pool related to the problem, then added to the existing optimized circuit, followed by parameter optimization for the newly obtained entire circuit. In the following, the algorithm repeats the process of circuit expansion based on the newly optimized circuit until the objective function converges. Although existing adaptive algorithms have effectively reduced redundant gates in VQE and QAOA quantum circuits, they still have some limitations. First, the evaluation methods typically rely on a single criterion mainly based on the gradient of a specific set of parameters, which might not fully capture the broader set of factors required for optimization. Secondly, \textcolor{black}{the existing adaptive algorithms only select one operator from the operator pool at a time, and then the entire circuit is optimized after each single addition}, which may require multiple rounds of optimization per run, thus increasing the number of iterations.

%第5段：介绍AMA-QAOA+算法提出的初衷，创新点
\medskip
In this paper, we propose an adaptive mixer allocation strategy tailored for QAOA+, referred to as the AMA-QAOA+ algorithm. AMA-QAOA+ also iteratively grows the ansatz circuit, initializing from a shallow circuit. However, unlike existing adaptive algorithms, \textcolor{black}{the mixer operator is selected} from a predefined operator pool based on an evaluation function that incorporates both the average gradient and the average initial expectation value over multiple sets of random parameters \textcolor{black}{in the AMA-QAOA+ algorithm}. This approach is to mitigate inefficient circuit designs resulting from reliance on a set of parameters or a single evaluation criterion. Additionally, AMA-QAOA+ performs optimization only after a certain number of \textcolor{black}{mixer operators} have been \textcolor{black}{sequentially selected and} added to minimize the number of iterations consumed in an optimization run. By integrating the adaptive idea, AMA-QAOA+ applies mixers to only a subset of qubits in each layer of the mixer unitary operator, and the set of qubits acted upon by the mixers tends to vary in each mixer layer. We focus on the maximum independent set (MIS) problem to quantify its performance. \textcolor{black}{Numerical simulations show that AMA-QAOA+ tends to achieve higher solution quality under the same number of optimization runs, along with requiring fewer CNOT gates} compared with original QAOA+, \textcolor{black}{the QAOA+ algorithm using the existing adaptive strategy (i.e., Adaptive-QAOA+), and another variant of QAOA+ that applies mixers to a randomly selected subset of qubits in each layer of the mixer unitary operator, and these qubits are non-uniform across different mixer layers (i.e., PNU) \cite{QLS}}. The detailed resource savings of AMA-QAOA+ in an optimization run are as follows.

\begin{itemize}
	\item \textcolor{black}{On \textcolor{black}{Erdős–Rényi (ER)} random graphs of an edge probability of 0.5, AMA-QAOA+ reduces the number of CNOT gates by 84.70\% and 76.71\% \textcolor{black}{compared with} QAOA+ and PNU, respectively. In comparison to Adaptive-QAOA+, it achieves a 14.61\% reduction in CNOT gates and a 48.90\% reduction in the number of iterations.}
	
	\item \textcolor{black}{On 3-regular graphs, AMA-QAOA+ reduces the number of CNOT gates by 74.82\% and 57.37\% \textcolor{black}{compared with} QAOA+ and PNU, respectively. In comparison to Adaptive-QAOA+, it achieves a 35.74\% reduction in the number of iterations.}
\end{itemize}

%\begin{itemize}
%\item On ER graphs, AMA-QAOA+ reduces the number of iterations and CNOT gates by $65.9711\%$ and $95.4306\%$, respectively, compared with QAOA+. Compared with PNU and Adaptive-QAOA+, AMA-QAOA+ achieves a reduction of $85.1878\%$ and $85.5850\%$ in the number of iterations, and $96.5394\%$ and $70.4889\%$ in the number of CNOT gates.  
%\item On 3-regular graphs, AMA-QAOA+ reduces the number of iterations and CNOT gates by $31.7478\%$ and $73.9322\%$, respectively, compared with QAOA+. Compared with PNU and Adaptive-QAOA+, AMA-QAOA+ achieves a reduction of $62.8483\%$ and $70.1106\%$ in the number of iterations, and $79.2711\%$ and $27.1683\%$ in the number of CNOT gates.
%\end{itemize}

\medskip
The paper is organized as follows. Section~\ref{preli} introduces the MIS problem and reviews the original QAOA and QAOA+ algorithms. Section~\ref{AMA} offers a detailed explanation of the AMA-QAOA+algorithm, including its key concepts and process. In Section~\ref{simulation}, we present the performance comparison results of different algorithms. Finally, Section~\ref{conclusion} summarizes the main contributions of this paper and discusses potential directions for future research.

\section{Preliminaries} \label{preli}
In this section, we review some relevant preliminary work to help readers better understand our work.

\subsection{The Maximum Independent Set Problem}

The MIS problem is an important CCOP that arises in various fields, such as network design \cite{MIS_in_network} and scheduling \cite{MIS_in_scheduling}, \textcolor{black}{and it is} equivalent to the minimum vertex cover or finding a maximum clique of the complementary graph \cite{MIS_vertex_cover}. Given an undirected graph \( G = (V, E) \), where \( V = \{1, 2, \dots, n\} \) represents the set of vertices and \( E = \{\{u, v\} \mid u, v \in V, u \neq v\} \) denotes the set of edges. In graph theory, two vertices \( u \) and \( v \) are said to be adjacent if there exists an edge between them, i.e., \( \{u, v\} \in E \). A subset of vertices \( V_s \subseteq V \) is called an independent set if no two vertices in \( V_s \) are adjacent. In other words, for every pair of vertices \( u, v \in V_s \), there is no edge between them, i.e., \( \{u, v\} \notin E \). A maximum independent set is an independent set that contains the maximal number of vertices among all independent sets in the graph \cite{MIS_definition}.

\medskip
To formalize the MIS problem, a binary variable \( x_v \) for each vertex \( v \in V \) is introduced \cite{QAOA,MLI}, where \( x_v = 1 \) if \( v \) belongs to the vertex subset \( V_s \), and \( x_v = 0 \) otherwise. A particular configuration of the vertices in the independent set can thus be represented by a bitstring \( x = x_1 x_2 \dots x_n \), where each bit corresponds to whether a vertex is included in the independent set or not. There are \( 2^n \) possible ways to assign values to the bits in \( x \).

\medskip
The objective of the MIS problem is to find a subset \( V_s \) that maximizes the number of vertices in the independent set, which is captured by the classical objective function
\vspace{-0.5em}
\begin{equation}
	C(x) = \sum_{v=1}^{n} x_v,
	\label{eq:max_independent_set}
	\vspace{-0.5em}
\end{equation}
subject to the independence constraint \( \sum_{(u, v) \in E} x_u x_v = 0 \), ensuring that no two vertices in \( V_s \) are adjacent. This constraint can be incorporated into the objective function using the Lagrange multiplier method, resulting in the following classical Hamiltonian \cite{QIRO}
\vspace{-0.5em}
\begin{equation}
	H_\text{cla} (x) = \sum_{v=1}^{n} x_v - \lambda \sum_{(u,v) \in E} x_u x_v,
	\label{eq:H_C}
	\vspace{-0.5em}
\end{equation}
where \( \lambda > 1 \) is the Lagrange multiplier that enforces the independence constraint \cite{Ising}. The first term in \( H_\text{cla} \) represents the objective to be maximized, while the second term penalizes the selection of adjacent vertices, ensuring that the solution respects the independence constraint. By maximizing this classical Hamiltonian, the optimal solution to the MIS problem can be found. That is, the maximum of $H_{\text{cla}}(x)$ equals the number of vertices in the MIS solution.

\medskip
However, the MIS problem is NP-Hard \cite{MIS_definition}, meaning that it is computationally intractable for large instances, as no known algorithm can solve it in polynomial time. As the number of vertices in the graph increases, the time complexity of solving the MIS problem grows exponentially, making it increasingly difficult for conventional algorithms, such as brute-force or exact optimization methods. To handle large-scale instances, approximate algorithms or heuristic methods are commonly employed to solve the MIS problem, offering a trade-off between accuracy and computational efficiency.

\medskip
%\textcolor{black}{The MIS problem is proven to be NP-hard. Current research on the MIS problem mainly focuses on two types of algorithms, one is exact algorithms \cite{exact_algorithm_MIS}, such as backtracking \cite{backtracking_algorithm_MIS} and branch-and-bound algorithms \cite{branch_and_bound_MIS}. These algorithms guarantee the optimal solution but usually have high computational complexity, making them suitable for small-scale problems. The other type is heuristic algorithms, such as local search \cite{local_search_MIS} and evolutionary algorithms \cite{evolutionary_algorithm_MIS}. These algorithms aim to provide quasi-optimal solutions and are more suitable for large-scale problems.}

\subsection{Quantum Approximate Optimization Algorithm}
QAOA is an approximate algorithm and provides an approximate solution to the problem. In the QAOA framework, the MIS problem is encoded in the ground state (i.e., the minimal energy state) of a quantum target Hamiltonian \( H_C \). This encoding is achieved by converting each binary variable \( x_{v} \) in $-H_\text{cla}(x)$ to the operator \( \frac{I - \sigma_v^z}{2} \) \cite{QAOA}, where \( \sigma_v^z \) denotes acting the Pauli-Z operator on the \( v \)-th qubit. The specific formulation of the quantum target Hamiltonian for the MIS problem in QAOA is given by
\vspace{-0.35em}
\begin{equation}
	H_C = - \sum_{v=1}^{n} \left( \frac{I - \sigma_v^z}{2} \right) + \lambda \sum_{(u, v) \in E} \left( \frac{I - \sigma_u^z}{2} \right) \left( \frac{I - \sigma_v^z}{2} \right).
	\label{eq:qaoa_HC}
\end{equation}

To achieve the MIS solution, QAOA starts with the ground state \( |s\rangle \) of the mixer quantum Hamiltonian \( H_M \) and evolves toward the target ground state through a PQC constructed using a \( p \)-layer QAOA ansatz, where \( p \) is referred to as the layer depth, and the initial state \( |s\rangle \) is required to be easily implementable, meaning it can be prepared by a constant-depth (i.e., the circuit does not scale with the problem size) quantum circuit starting from the $|0 \rangle ^{\otimes n } $ state.

\medskip
Each layer of the QAOA ansatz consists of two unitary operators $\mathrm{e}^{- \mathrm{i}\gamma_i H_C}$ and $\mathrm{e}^{- \mathrm{i}\beta_i H_M}$, where $\gamma_i$ and $\beta_i$ are the circuit parameters in the $i$-th layer of the QAOA ansatz, and $\mathrm{i}$ denotes the imaginary unit. In QAOA, the mixer Hamiltonian is typically chosen as $H_M = -\sum_{j=1}^n \sigma_j^x$ because its ground state $|s\rangle = |+\rangle^{\otimes n}$ can be prepared using a depth-1 quantum circuit by applying Hadamard gates to each qubit in the $|0 \rangle ^{\otimes n } $ state, where $\sigma_j^x$ represents acting the Pauli-X operator on the $j$-th qubit. The system evolves towards the ground state of the target Hamiltonian \( H_C \) through a PQC with the evolution controlled by the parameters \( \gamma_i \) and \( \beta_i \).

\medskip
The quantum state $|s\rangle$ evolves through $p$-layer QAOA ansatz, resulting in the final quantum state
\vspace{-0.35em}
\begin{equation}
	|\boldsymbol{\gamma}_p, \boldsymbol{\beta}_p\rangle = \prod_{i=1}^p \mathrm{e}^{- \mathrm{i}\beta_i H_M} \mathrm{e}^{- \mathrm{i}\gamma_i H_C} |s\rangle,
	\label{eq:output_state}
	\vspace{-0.5em}
\end{equation}
where $\boldsymbol{\gamma}_p = (\gamma_{1}, \gamma_{2}, \dots, \gamma_{p})$ and $\boldsymbol{\beta}_p = (\beta_{1}, \beta_{2}, \dots, \beta_{p})$ are the variational parameters in the PQC, \textcolor{black}{and the parameters \(\gamma_i\) and \(\beta_i\) are respectively within the range \([0, 2\pi]\) and \([0, \pi]\) \cite{QAOA}}. The average expectation function value of the output state with respect to the target Hamiltonian is given by
\vspace{-0.35em}
\begin{equation}
	F(\boldsymbol{\gamma}_p, \boldsymbol{\beta}_p) = \langle \boldsymbol{\gamma}_p, \boldsymbol{\beta}_p | H_C | \boldsymbol{\gamma}_p, \boldsymbol{\beta}_p \rangle,
	\label{eq:cost_function}
	\vspace{-0.5em}
\end{equation}
\textcolor{black}{which can be calculated by repeated measurements of the output quantum state using the computational basis.} The target ground state corresponds to the minimum expectation function value. Therefore, by minimizing this expectation function, QAOA can find the quantum state that encodes the optimal solution to the problem.

\medskip
Numerous optimization runs are required to find the optimal parameters that minimize the expectation function. In each run, an initial set of parameters is assigned to the circuit, and the quantum state evolves according to the quantum circuit. The quantum computer then performs measurements to compute the expectation value of the target Hamiltonian. These results are passed to the classical optimizer, which updates the QAOA parameters \( (\boldsymbol{\gamma}_p, \boldsymbol{\beta}_p) \). The updated parameters are sent back to the quantum computer for the next round of evolution and calculation. This process repeats until a stopping condition is met, such as convergence or reaching the maximum number of iterations \cite{poisson, MLI}.

\medskip
To quantify the quality of the final quantum state, the approximation ratio (AR) 
\vspace{-0.35em}
\begin{equation}
	r = \frac{F(\boldsymbol{\gamma}_p, \boldsymbol{\beta}_p)}{F_{\text{min}}},
	\label{AR}
	\vspace{-0.5em}
\end{equation}
is introduced \cite{QAOA,MLI,TQA}, where \( F_{\text{min}} \) is the ground state energy of \( H_C \), \textcolor{black}{which equals to the minimum value of $-H_{\text{cla}}(x)$}. The AR measures how close the output state provided by QAOA is to the ground state of \( H_C \), with \( r \le 1 \), where a value of 1 indicates the optimal solution. \textcolor{black}{In this paper, to ensure an accurate calculation of the approximation ratio, we obtain the minimum value of $-H_{\text{cla}}(x)$ through the exhaustive search for each graph instance used in our numerical simulations.}

\begin{figure*}[htbp]
	\centering
	% 子图1
	\begin{minipage}[b]{0.32\textwidth} % 设置子图宽度为总宽度的45%
		\centering
		\subfloat[ A given ER graph.]{\label{random_ER} \includegraphics[width=\textwidth]{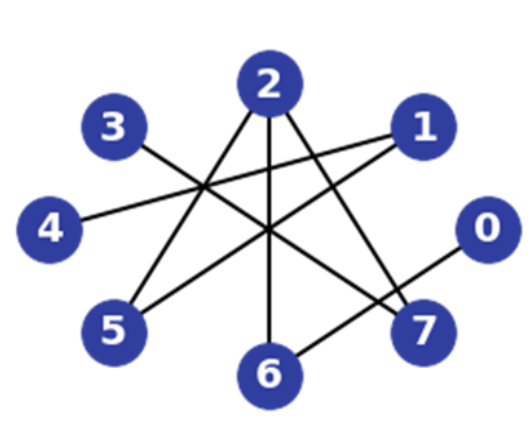}} % 替换为子图的文件名
	\end{minipage}
	\hspace{0.02\textwidth} % 调节间距
	% 子图2
	\begin{minipage}[b]{0.465\textwidth} % 设置子图宽度为总宽度的45%
		\centering
		\subfloat[ One layer QAOA+ ansatz.]{\label{original_qaoa+} \includegraphics[width=\textwidth]{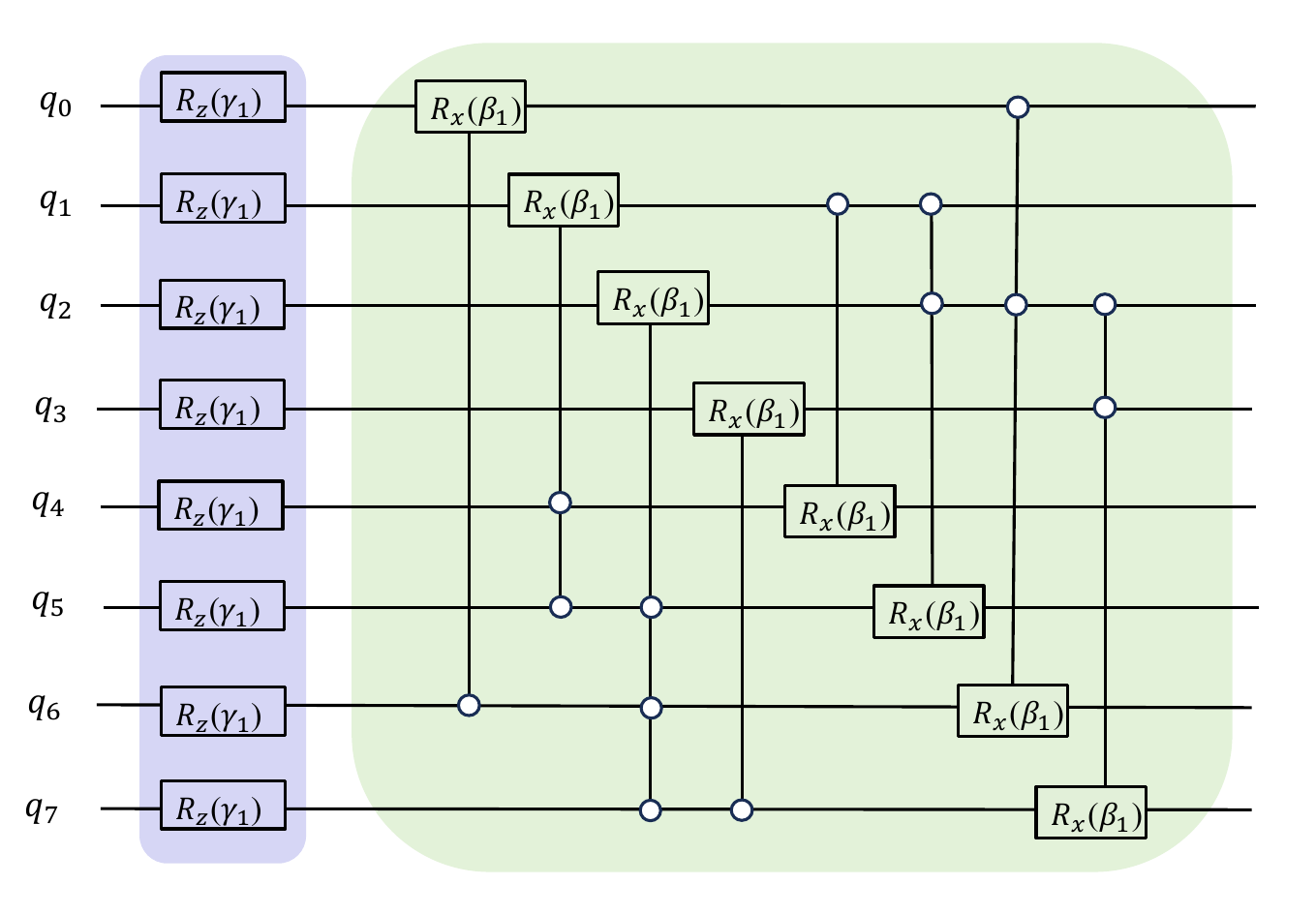}} % 替换为子图的文件名
	\end{minipage}
	\captionsetup{justification=raggedright}  % 左对齐
	\setlength{\belowcaptionskip}{-0.45 cm}   %调整图片标题与下文距离
	\caption{The implementation of a single-layer QAOA+ ansatz for the given ER graph. (a) An ER graph with $n = 8$ vertices. (b) The circuits with the green and purple undertone correspond to $\mathrm{e}^{-\mathrm{i}\beta_{1}H_{M}}$ and $\mathrm{e}^{-\mathrm{i}\gamma_{1}H_{C}}$, respectively. Each qubit $q_v$ corresponds to vertex $v$, where $v = 0, 1, \dots, n-1$.} % 整体图的标题
	\label{QAOA+_instance} % 整体图的标签
\end{figure*}

\subsection{QAOA+}
When solving the CCOPs using QAOA, the Hilbert space encompasses both feasible solutions and infeasible solutions. This leads to a large search space, and the algorithm may yield infeasible solutions upon termination. To address these limitations of QAOA, Hadfield et al. \cite{QAOA+} proposed a novel quantum alternating operator ansatz algorithm, referred to as QAOA+. QAOA+ encodes the problem constraints into the mixer Hamiltonian $H_{M}$ and constructs a feasible space that only contains feasible solutions by alternatively applying $p$-layer QAOA+ ansatz on the initial quantum state $|s\rangle$, where $|s\rangle$ is a feasible state that is easy to create \cite{QAOA+}. By searching for the optimal solution within a feasible space, the size of the search space is reduced, and the probability of QAOA+ yielding invalid solutions is zero. Similar to the QAOA ansatz, each layer of the QAOA+ ansatz also consists of two unitary operators $\mathrm{e}^{- \mathrm{i}\gamma_i H_C}$ and $\mathrm{e}^{- \mathrm{i}\beta_i H_M}$, but the specific implementation of the Hamiltonian in QAOA+ differs from QAOA. In the following, for the MIS problem, the designs of $H_M$ and $H_C$ in QAOA+ are introduced. 

\medskip
In QAOA+, the mixer unitary operator is required to preserve and explore the feasible solution space. Exploration refers to the evolution of the state from one feasible quantum state (i.e., a single feasible solution or a superposition of multiple feasible solutions) to another feasible state. Preservation ensures that no infeasible quantum states are generated during the evolution process. For the MIS problem, the quantum unitary operations acting on feasible states can be understood as analogous to classical operations, such as the removal or addition of vertices, on independent sets. To clarify this process, we analyze the conditions that must be satisfied when performing a classical operation on an independent set to generate a new independent set.

\medskip
Denote the independent set as $V'$, and let the binary variable $x_v = 1$ indicate that vertex $v$ belongs to $V'$ and $x_v = 0$ otherwise. To add or remove vertices from $V'$ without breaking its independence, the following conditions must be satisfied.
\begin{enumerate}
	\item \textbf{Removing vertices from an independent set requires no additional conditions}. This is because removing vertices does not affect the independence of the remaining vertices and can result in another independent set. Successfully removing vertex $v$ from $V'$ corresponds to the state transition of $x_v$ from $1$ to $0$.
	
	\item \textbf{Adding a new vertex $v_\text{new}$ to $V'$ without destroying the independence constraint if and only if all its neighbors are not in $V'$ (i.e., all neighboring vertices of $v_\text{new}$  are in state $0$).} Here, $v_\text{new} \notin V'$ and $v_\text{new} \in V$. Successfully adding vertex $v_\text{new}$ into $V'$ corresponds to the state transition of $x_{v_\text{new}}$ from $0$ to $1$.
\end{enumerate}

\medskip
All in all, without destroying the independence, the addition or removal of a vertex $v$ can be controlled by the states of its neighbors. Specifically, the condition for adding or removing vertex $v$ while maintaining the independence of the set is that all neighbors of $v$ must not be in the independent set $V'$. This condition can be expressed as
\vspace{-0.5em}
\begin{equation}
	f_v = \prod_{v_{q_j} \in N(v)} (1 - x_{v_{q_j}}) = 1,
	\label{eq:neighbor_condition}
	\vspace{-0.5em}
\end{equation}
where $N(v)$ denotes the set of neighbors of vertex $v$ that contains $d(v)$ vertices, and $x_{v_{q_j}}$ is the state of the $q_j$-th neighbor of $v$. The condition $f_v = 1$ ensures that all neighbors of vertex $v$ are not in the independent set $V'$, thus maintaining the independence of the set when $v$ is added or removed.

\begin{figure*}[htbp]
	\centering
	% 子图1
	\begin{minipage}[b]{0.48\textwidth} % 设置子图宽度为总宽度的45%
		\centering
		\subfloat[ The circuit architecture of the QAOA+ algorithm.]{\label{original_qaoa} \includegraphics[width=\textwidth]{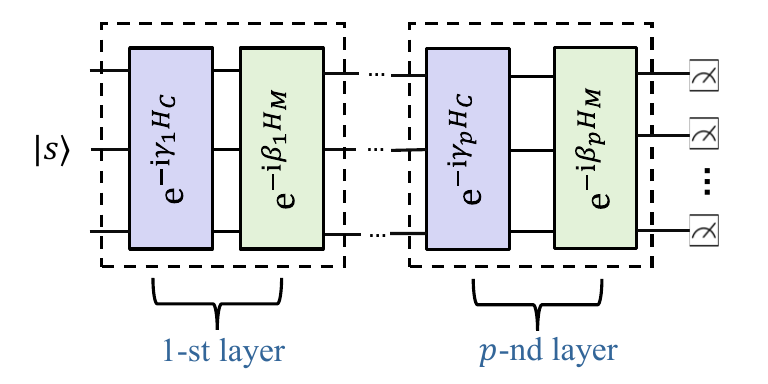}} % 替换为子图的文件名
	\end{minipage}
	\hspace{0.02\textwidth} % 调节间距
	% 子图2
	\begin{minipage}[b]{0.48\textwidth} % 设置子图宽度为总宽度的45%
		\centering
		\subfloat[ The circuit architecture of the AMA-QAOA+ algorithm.]{\label{adaptive_qaoa+} \includegraphics[width=\textwidth]{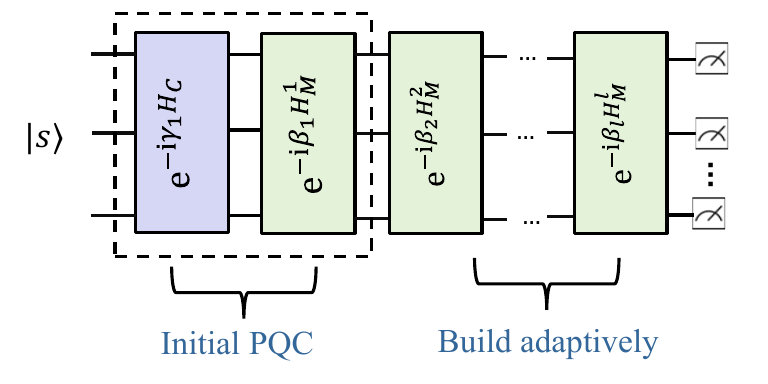}} % 替换为子图的文件名
	\end{minipage}
	\captionsetup{justification=raggedright}  % 左对齐
	\setlength{\belowcaptionskip}{-0.45 cm}   %调整图片标题与下文距离
	\caption{The quantum circuit architectures of QAOA+ and AMA-QAOA+, where $H_C = -\sum_{v=1}^{n} \frac{I - \sigma_v^z}{2}$. (a) In QAOA+, the PQC is built by $p$-layer QAOA+ ansatz, where one layer of QAOA+ ansatz has two unitary operators, and there are two optimized parameters in each layer. Notably, each qubit is acted on by the mixer when implementing each layer of the mixer unitary operator. (b) In the AMA-QAOA+ algorithm, an initial PQC is first built by unitary operators $U_{C}(\gamma_{1})$ and $U_{M}(\beta_{1})$. In the subsequent layers, there is only one layer of mixer unitary operation $U_{M}(\beta_{l})$. For $l \ge 1$, several mixers are randomly or adaptively added into $U_{M}(\beta_{l})$.} % 整体图的标题
	\label{PQC} % 整体图的标签
\end{figure*}

\medskip
Successfully removing a vertex from $V'$ or adding a vertex to $V'$ corresponds to applying the $X$ gate to the qubit corresponding to vertex $v$. If the operation fails (i.e., a vertex is not successfully removed or added), the identity operation $I$ is applied instead. For each vertex $v$, its corresponding partial mixer Hamiltonian can be formally expressed as 
\vspace{-0.5em}
\begin{equation}
	\begin{aligned}
		M_v = & \sum_{f_v=1} \left| x_{v_{q_0}} x_{v_{q_1}} \dots x_{v_{q_d(v)}} \right\rangle 
		\left\langle x_{v_{q_0}} x_{v_{q_1}} \dots x_{v_{q_d(v)}} \right| \otimes X_v \\
		&+ \sum_{f_v=0} \left| x_{v_{q_0}} x_{v_{q_1}} \dots x_{v_{q_d(v)}} \right\rangle 
		\left\langle x_{v_{q_0}} x_{v_{q_1}} \dots x_{v_{q_d(v)}} \right| \otimes I_v.
	\end{aligned}
	\label{eq:mixer_operator}
	\vspace{-0.5em}
\end{equation}
In QAOA+, the total mixer Hamiltonian is given by $H_M = \sum_{v=1}^{n} M_v$, and the mixer unitary operator, parameterized by $\beta_i$, is expressed as 
\begin{equation}
	U_M(\beta_i ) = \mathrm{e}^{-\mathrm{i}\beta_i H_M}.
	\label{eq:mixer_operator}
\end{equation}

For any vertex $v$ with a degree greater than one, the implementation of \textcolor{black}{$U_{M_{v} }(\beta_{i}) =$ } $\mathrm{e}^{-\mathrm{i}\beta_i M_v}$ corresponds to a multi-qubit controlled $R_x(\beta_{i})$ gate \textcolor{black}{(i.e., the `mixer')}, where the relevant qubits corresponding to the neighbors of $v$ act as control qubits and the qubit corresponding to $v$ serves as the target qubit. This implies that each vertex (represented by a qubit) has an associated multi-qubit controlled $R_x$ gate. Since current quantum hardware is generally unable to directly implement multi-qubit controlled $R_x$ gates, it is necessary to decompose them into a series of basic gates (e.g., controlled-NOT gates and rotation gates). As the number of control qubits increases, the resources and circuit depth required for this decomposition grow dramatically \cite{gate_decomposition_ieee,gate_decomposition_quantum}, which limits the scale of optimization problems that can be handled by resource-constrained devices. 
%Therefore, when solving the MIS problem using QAOA+, it is crucial to minimize not only the total number of multi-qubit controlled gates but also to prioritize reducing those with more control qubits. 

\medskip
QAOA+ searches for the optimal solution within the feasible solution space, aiming to maximize the number of vertices in the independent set. The corresponding classical objective function is given in \textbf{Equation~\eqref{eq:max_independent_set}}. To express this objective in the quantum domain, each $x_v$ in the classical cost function $-C(x)$ is replaced by the quantum operator $(I - \sigma_v^z)/2$ \cite{QAOA}. The quantum target Hamiltonian for the MIS problem in QAOA+ is given by
\vspace{-0.5em}
\begin{equation}
	H_C = -\sum_{v=1}^{n} \frac{I - \sigma_v^z}{2}.
	\label{eq:cost_hamiltonian}
	\vspace{-0.25em}
\end{equation}

The corresponding target unitary operator for this Hamiltonian, parameterized by $\gamma_i$, is expressed as
\vspace{-0.5em}
\begin{equation}
	U_C(\gamma_i) = \mathrm{e}^{-\mathrm{i}\gamma_i H_C} = \prod_{v=1}^{n} R_z(\gamma_i, v),
	\label{eq:unitary_operator}
	\vspace{-0.5em}
\end{equation}
which is composed of a sequence of single-qubit $R_z$ rotation gates. These $R_z(\gamma_i, v)$ gates apply a phase shift to the quantum state of qubit $v$, determined by the parameter $\gamma_i$. Importantly, while the $R_z$ rotations modify the relative phase of the quantum state, they do not flip the individual qubits between the \( |0\rangle \) and \( |1\rangle \) states. As a result, the $R_z$ operator only alters the amplitude (or phase) of the quantum state, rather than causing qubit-flip operations. For a better understanding of the PQC of the QAOA+, we give the implementation of a single-layer QAOA+ ansatz for a given ER graph in \textbf{Figure~\ref{QAOA+_instance}}  and the circuit architecture of QAOA+ in \textbf{Figure~\ref{original_qaoa}}.

\medskip
In summary, QAOA+ improves upon QAOA by encoding problem constraints into the mixer Hamiltonian, restricting the search to only feasible solutions. This effectively reduces unnecessary exploration. However, in the QAOA+, mixers are applied to each qubit in each layer of the mixer unitary operator, which may result in high gate overheads. To reduce the number of mixers required by QAOA+ for solving CCOPs, the AMA-QAOA+ algorithm is proposed.

\section{The Adaptive Mixer Allocation strategy for the QAOA+} \label{AMA}
%In the current NISQ devices, the limitations of quantum gate resources directly determine the feasibility of algorithms and the problem sizes supported by the hardware. In particular, multi-qubit gates have significantly higher error rates compared with single-qubit gates. As the number of multi-qubit gates increases, noise accumulation severely affects the fidelity of quantum states, thereby reducing the solution quality. Moreover, an increase in circuit depth poses serious hardware bottlenecks. Due to the finite coherence time of qubits in current quantum devices, excessive circuit depth can cause quantum states to lose their validity through decoherence before the evolution is completed. Therefore, reducing the number of multi-qubit gates and circuit depth is crucial for algorithm performance in NISQ devices. 
This section introduces the purpose and basic framework of the AMA-QAOA+ algorithm.

\medskip
AMA-QAOA+ aims to construct a feasible solution space by adaptively applying the mixers to a subset of qubits in each layer of the mixer unitary operation, and then it searches for the solution within this space. Notably, AMA-QAOA+ allows the qubits acted on by mixers to vary in each layer of the mixer unitary operator. To achieve the above goal, AMA-QAOA+ starts with an initial pre-trained PQC and incrementally adds a new layer of mixer unitary operator to the trained circuit. In each layer of the mixer unitary operator, these qubits acted upon by the mixer are adaptively selected according to an evaluation function, and we will return shortly to the question of how to design the function in the following content. After adding a new layer of mixer unitary operator to the trained circuit, AMA-QAOA+ optimizes the newly obtained entire circuit \textcolor{black}{ until the absolute change in the expectation value is less than 0.01 over three successive iterations \cite{MLI}, and this process is a round of optimization}. Subsequently, AMA-QAOA+ repeats the process of adding the new layer of mixer unitary operation and performs parameter optimization based on the newly trained circuit until a predefined stopping condition 
\begin{equation}
	\textcolor{black}{|F_t - F_{t-1}| \le \delta \quad \text{and} \quad |F_{t-1} - F_{t-2}| \le \delta}
	\label{eq:stopping_condition}
\end{equation}
is satisfied, \textcolor{black}{ where \( F_t \) denotes the expectation function value of the output quantum state with respect to the target Hamiltonian \( H_C \) after the \( t \)-th round of parameter optimization, and \( \delta \) is a threshold that judges whether the expectation function reaches convergence.} \textcolor{black}{This stopping criterion is designed to timely detect the convergence as well as prevent premature termination due to insufficient stabilization of the expectation function. Specifically, if convergence is determined based on only two consecutive rounds, the algorithm may terminate without truly reaching a stable solution. Conversely, requiring four or more rounds may introduce unnecessary computational overhead while providing limited improvement in convergence reliability.}

%A smaller \( \delta \) leads to stricter convergence criteria and more optimization rounds, which may improve solution quality at the cost of higher computational overhead, while a larger \( \delta \) results in earlier termination, which may lead to suboptimal solutions due to insufficient convergence.

\medskip
Notably, the target unitary operator primarily influences the amplitude of the quantum state without directly affecting the flip of individual qubits. In contrast, the mixer unitary operator plays a more crucial role in each layer by influencing qubit flips, thereby facilitating the search process. \textcolor{black}{To reduce the number of optimized parameters and simplify the optimization process, we discard the target unitary operator after the first layer of ansatz, thereby breaking the alternating structure of the standard QAOA+, inspired by Ref.~\cite{UFLP}. This design choice allows us to focus optimization on the more impactful mixer layers. In \textbf{Appendix~\ref{com}}, we further investigate the performance of the adaptive algorithm in two settings, that is, one retains the alternating structure and one discards it after the first layer of the ansatz.  Numerical results confirm that for AMA-QAOA+, removing the alternating structure leads to faster convergence and higher-quality solutions.}
%Based on this property, we argue that repeatedly applying the target \textcolor{black}{unitary circuit} in each layer may introduce redundant parameters without significantly improving the search capability.

\medskip
All in all, AMA-QAOA+ consists of three main steps, that is, constructing the initial pre-trained PQC, adaptively adding partial mixers for the new layer of mixer unitary operator, and optimizing the entire circuit. The circuit architecture of AMA-QAOA+ is given in \textbf{Figure~\ref{adaptive_qaoa+}}. The details of each step are as follows.

\subsection{The construction of the initial pre-trained PQC}
In the AMA-QAOA+ algorithm, the initial pre-trained PQC is first constructed using a single layer of the QAOA+ ansatz, corresponding to the dashed box in Figure~\ref{adaptive_qaoa+}. During the implementation of the mixer unitary operation, to reduce the number of mixers, only a subset of qubits is randomly selected for applying the mixers. Given that the implementation cost of single-qubit rotation gates \( R_z \) is relatively low, these \( R_z \) gates are applied to all qubits when constructing the target unitary operator. Once the initial circuit is built, its parameters are randomly initialized. The parameters are subsequently optimized using a classical optimizer, yielding the pre-trained PQC, denoted as \( U_{\text{init}}(\gamma_{1}^{*},\beta_{1}^{*}) \), where $\gamma_{1}^{*}$ and $\beta_{1}^{*}$ are the relevant optimized parameters of the first ansatz.

\subsection{The adaptive mixer allocation strategy}
%Unlike existing methods, which rely on the gradient values under a fixed set of circuit parameters to decide which gate to select from the operator pool, the evaluation function values under multiple sets of initial parameters are taken into account, which is to reduce the potential randomness introduced by the initial parameter selection. Furthermore, by considering both the gradient and the initial expectation value, rather than relying solely on the initial gradient information, our method may achieve higher-quality solutions during the optimization process.

Since each qubit $j$ is associated with a unique mixer operator \textcolor{black}{$U_{M_{j} }(\beta_{i}) = \mathrm{e}^{-\mathrm{i}\beta_i M_j}$}, selecting which qubits to apply the mixers to is equivalent to selecting which mixer operators from the operator pool \( \textcolor{black}{\mathcal{U}_{\text{pool}} } = \{ \textcolor{black}{U_{M_j}(\beta_{i})} \,|\, j = 1, 2, \dots, n \} \) to be added into the $i(\ge 2)$-th layer of the \textcolor{black}{mixer unitary operator}. In this work, an evaluation function is used to guide the selection of mixer operators from the pool. Specifically, the evaluation function \textcolor{black}{with respect to the $U_{M_{j}}(\beta_{i})$} is defined as
\begin{equation}
	\begin{split}
		C(U_{M_j})  &=  (1 - f_1)\times F_{\text{fun}}(\gamma_{1}^{*}, \boldsymbol{\beta}_{1:i-1}^{*}, U_{M_j}(\beta_{i})) \\
		& \quad + f_1 \times \textcolor{black}{F_{\text{grad}}(\gamma_{1}^{*}, \boldsymbol{\beta}_{1:i-1}^{*}, U_{M_j}(\beta_{i}))}
	\end{split}
	\label{evaluation_function}
\end{equation}
\textcolor{black}{where $\boldsymbol{\beta}_{1:i-1}^{*}$ denotes the optimized parameters from the first to \((i-1)\)-th mixer layers, and \( f_1 \in (0, 1) \) is a weighting coefficient that adjusts the contribution of the \( F_{\mathrm{fun}} \) and \( F_{\mathrm{grad}} \).}

\medskip
\textcolor{black}{
	\( F_{\mathrm{fun}} \) represents the average initial expectation function value of the output state with respect to the target Hamiltonian \( H_C \), evaluated over \( N (>1) \) sets of randomly initialized parameters. Specifically, }
\begin{equation}
	\textcolor{black}{
		F_{\text{fun}}(\gamma_{1}^{*}, \boldsymbol{\beta}_{1:i-1}^{*}, U_{M_j}(\beta_{i})) = \frac{-1}{N} \sum_{k=1}^{N} \textcolor{black}{F_{\text{init}}}(\gamma_{1}^{*}, \boldsymbol{\beta}_{1:i-1}^{*}, U_{M_j}(\beta_{i}^{\textcolor{black}{(k)}})),
	}
	\vspace{-0.5em}
	\label{eq:F_fun}
\end{equation}
\textcolor{black}{where each \textcolor{black}{\( F_{\text{init}} \)} corresponds to the expectation value of the output state with respect to the target Hamiltonian \( H_C \) under the \( k \)-th set of random parameters $(\gamma_{1}^{*}, \boldsymbol{\beta}_{1:i-1}^{*}, \beta_{i}^{\textcolor{black}{(k)}})$. More details of the calculation procedure for\textcolor{black}{\( F_{\text{init}} \)} are provided in \textbf{Appendix~\ref{calculate_F_grad}}. Here, we introduce the negative sign in \textbf{Equation~\eqref{eq:F_fun}} to ensure that a better initial state with respect to $H_{C}$ yields a higher initial value. $F_{\text{fun}}$ reflects the quality of the initial solution that may be achieved by adding the candidate mixer operator. A larger initial expectation value often indicates that the circuit with the added mixer tends to produce better-quality initial states even before optimization. Such mixers are more likely to accelerate convergence or help the algorithm escape poor local minima \cite{INTERP,TQA}. }

\medskip
\textcolor{black}{\( F_{\text{grad}} \) denotes the average gradient of the expectation function with respect to the newly introduced parameter \( \beta_i \), computed using the parameter-shift rule \cite{PSR}  over the same \( N \) sets of parameter initializations. Specifically,}
\begin{equation}
	\textcolor{black}{F_{\text{grad}}(\gamma_{1}^{*}, \boldsymbol{\beta}_{1:i-1}^{*}, U_{M_j}(\beta_{i})) = \frac{1}{N} \sum_{k=1}^{N} \left| \textcolor{black}{\frac{\partial F}{\partial \beta_i^{(k)}}} \right|,}
	\vspace{-0.5em}
\end{equation}
where \textcolor{black}{$\frac{\partial F}{\partial \beta_i^{(k)}}$} is the gradient of the expectation function with respect to \( \beta_i \) under $k$-th set of random initial parameters $(\gamma_{1}^{*}, \boldsymbol{\beta}_{1:i-1}^{*}, \beta_{i}^{\textcolor{black}{(k)}})$, and a detailed computation is also provided in Appendix~\ref{calculate_F_grad}. A larger average gradient indicates that the addition of the candidate mixer operator can induce more significant changes in the optimization landscape, thereby enhancing the circuit’s trainability and increasing the likelihood of escaping from local optima \cite{adaptive_VQE}.

\begin{algorithm}[H]
	\caption{AMA-QAOA+}
	\label{AMA_steps}

	\begin{algorithmic}[1]
		
		\Statex \textbf{\textcolor{black}{Input:}} \textcolor{black}{\( G = (V, E) \), thresholds \( \delta_{\text{grad}}, \delta_{\text{add}}, \delta \).}
		\Statex \textbf{\textcolor{black}{Output:}} \textcolor{black}{The constructed parameterized circuit and its output quantum state}

		\State \textbf{Step 1:} Construct the pre-trained PQC \( U_{\text{init}}(\gamma_{1}^{*},\beta_{1}^{*}) \) with a single layer of QAOA+ ansatz, and there are only partial qubits randomly selected for applying the mixers in this mixer layer.
		
		\State \textbf{Step 2:} \textcolor{black}{ Adaptively construct the $i(\ge 2)$-th layer of the mixer unitary operator.}
		
		\State \hspace{3mm} \textbf{Step 2.1:} Initialize the operator pool \( \textcolor{black}{\mathcal{U}_{\text{pool}} } = \{ \textcolor{black}{U_{M_j}(\beta_{i})} \,|\, j = 1, 2, \dots, n \} \), $M_{\text{add}} = 0$, \textcolor{black}{\( U_{\text{init}} = U_{\text{init}}(\gamma_{1}^{*},\boldsymbol{\beta}_{1:i-1}^{*}) \), where $\boldsymbol{\beta}_{1:i-1}^{*}$ denotes the optimized parameters from the first to \((i-1)\)-th mixer layers}. 
		
		\State \hspace{3mm} \textbf{Step 2.2:}  Adaptively select partial mixer operators from the pool $\textcolor{black}{\mathcal{U}_{\text{pool}} }$ and add them into the circuit \textcolor{black}{$U_{\text{init}}$.}
		
		\State \hspace{4mm}   \textbf{step 2.2.1} Compute the evaluation function value for each mixer operator in the current operator pool. The calculation process is as follows. 
		\begin{itemize}
			
			\item Select the operator \( \textcolor{black}{U_{M_j}(\beta_{i})} \) from the pool \( \textcolor{black}{\mathcal{U}_{\text{pool}} } \), and add it to the circuit $U_{\text{init}}$, obtaining the current circuit \textcolor{black}{$U_{\text{cur}}(\gamma_{1}^{*}, \boldsymbol{\beta}_{1:i-1}^{*},\beta_{i}) = U_{M_j}(\beta_{i}) \cdot U_{\text{init}}$}.
			
			\item Generate multiple sets of \textcolor{black}{random} parameters for the current circuit. Specifically, the parameter $\beta_{i}$ of the $i$-th layer of the mixer unitary operator is randomly generated, and the $U_{\text{init}}$ reuses the optimized parameters \textcolor{black}{$(\gamma_{1}^{*}, \boldsymbol{\beta}_{1:i-1}^{*})$}.
			
			\item Calculate the corresponding gradients (initial expectation function values) for the current circuit under different circuit parameters, then average the obtained gradients (initial expectation function values).
			
			\item Compute the corresponding evaluation function value with respect to the mixer operator \( \textcolor{black}{U_{M_j}(\beta_{i})} \).
		\end{itemize}
		
		\State \hspace{4mm} \textbf{step 2.2.2} Select one mixer operator that maximizes the evaluation function in the current operator pool, and this mixer operator is denoted as \( \textcolor{black}{U_{M_{j'}^{*}}(\beta_{i})} \).
		
		\State \hspace{3mm} \textbf{step 2.2.3} Update the pool $\textcolor{black}{\mathcal{U}_{\text{pool}} } = \textcolor{black}{\mathcal{U}_{\text{pool}} } \setminus \textcolor{black}{U_{M_{j'}^{*}}(\beta_{i})}$, $M_{\text{add}} = M_{\text{add}} + 1 $, and \textcolor{black}{$U_{\text{init}} = U_{\text{init}}(\gamma_{1}^{*}, \boldsymbol{\beta}_{1:i-1}^{*},\beta_{i}) = U_{M_{j'}^{*}}(\beta_{i}) \cdot U_{\text{init}}$}.

		\State \hspace{4mm} \textbf{step 2.2.4} Repeat \textbf{step 2.2.1-step 2.2.3} based on the newly updated operator pool, $M_{\text{add}}$ and $U_{\text{init}}$ when the maximum $\textcolor{black}{F_{\text{grad}}}$ \(  \ge \delta_{\text{grad}} \) and $M_{\text{add}}$ \(  <\delta_{\text{add}} \).
		
		\State \textbf{Step 3:}  Optimize the newly obtained entire circuit $U_{\text{init}}(\gamma_{1}^{*}, \boldsymbol{\beta}_{1:i-1}^{*},\beta_{i})$, \textcolor{black}{
			where the newly introduced parameter \( \beta_{i} \) is randomly initialized and then jointly optimized together with the previously optimized parameters \( \gamma_{1}^{*} \) and \( \boldsymbol{\beta}_{1:i-1}^{*} \). After the optimization completes, all parameters are updated, and the number of mixer layers is increased by one, i.e., \( i = i + 1 \)}.
		
		\While{stopping condition in Equation~\eqref{eq:stopping_condition} not met}
		\State Repeat \textbf{Step 2 - Step 3}.
		\EndWhile
	\end{algorithmic}
\end{algorithm}

\medskip
Based on the above evaluation function, AMA-QAOA+ sequentially selects multiple mixer operators from the predefined operator pool through an iterative selection process, and these selected operators construct a new layer of the mixer unitary operator. Here are the basic steps. (1) Initialize $U_{\text{init}} = U_{\text{init}}(\gamma_{1}^{*},\boldsymbol{\beta}_{1:i-1}^{*}) $. (2) For each mixer operator in \( \textcolor{black}{\mathcal{U}_{\text{pool}} } \), AMA-QAOA+ first computes its corresponding evaluation function value based on $U_{\text{init}}$, (3) and then selects the mixer operator \( \textcolor{black}{U_{M_{j'}^{*}}(\beta_{i})} \) that maximizes the evaluation function from the operator pool \( \textcolor{black}{\mathcal{U}_{\text{pool}} } \). (4) The selected mixer is added to the circuit $U_{\text{init}}$, resulting in a new circuit $U_{\text{init}} = U_{M_{j'}^{*}}(\beta_{i}) \cdot U_{\text{init}}$, and the operator pool is updated by removing the selected mixer operator. (5) The steps (2)-(4) are repeated based on the updated pool $\mathcal{U}_{\text{pool}}$ and $U_{\text{init}}$ until the number of added mixer operators \( M_{\text{add}} \) reach the upper limit \( \delta_{\text{add}} \) or the obtained maximal \( \textcolor{black}{F_{\text{grad}}}(\gamma_{1}^{*}, \boldsymbol{\beta}_{1:i-1}^{*}, U_{M_j}(\beta_{i})) \) based on the current operator pool falls below the threshold \( \delta_{\text{grad}} \). (5) Finally, the selected mixer operators construct the $i$-th layer of mixer unitary operations. \textcolor{black}{Notably, the detailed steps of computing $C(U_{M_{j}})$ are given in Algorithm~\ref{AMA_steps}, and we omit them to avoid repetition.}

\begin{figure*}[htbp]
	\centering
	% 子图1
	\begin{minipage}[b]{0.985\textwidth} % 设置子图宽度为总宽度的45%
		\centering
		\subfloat{ \includegraphics[width=\textwidth]{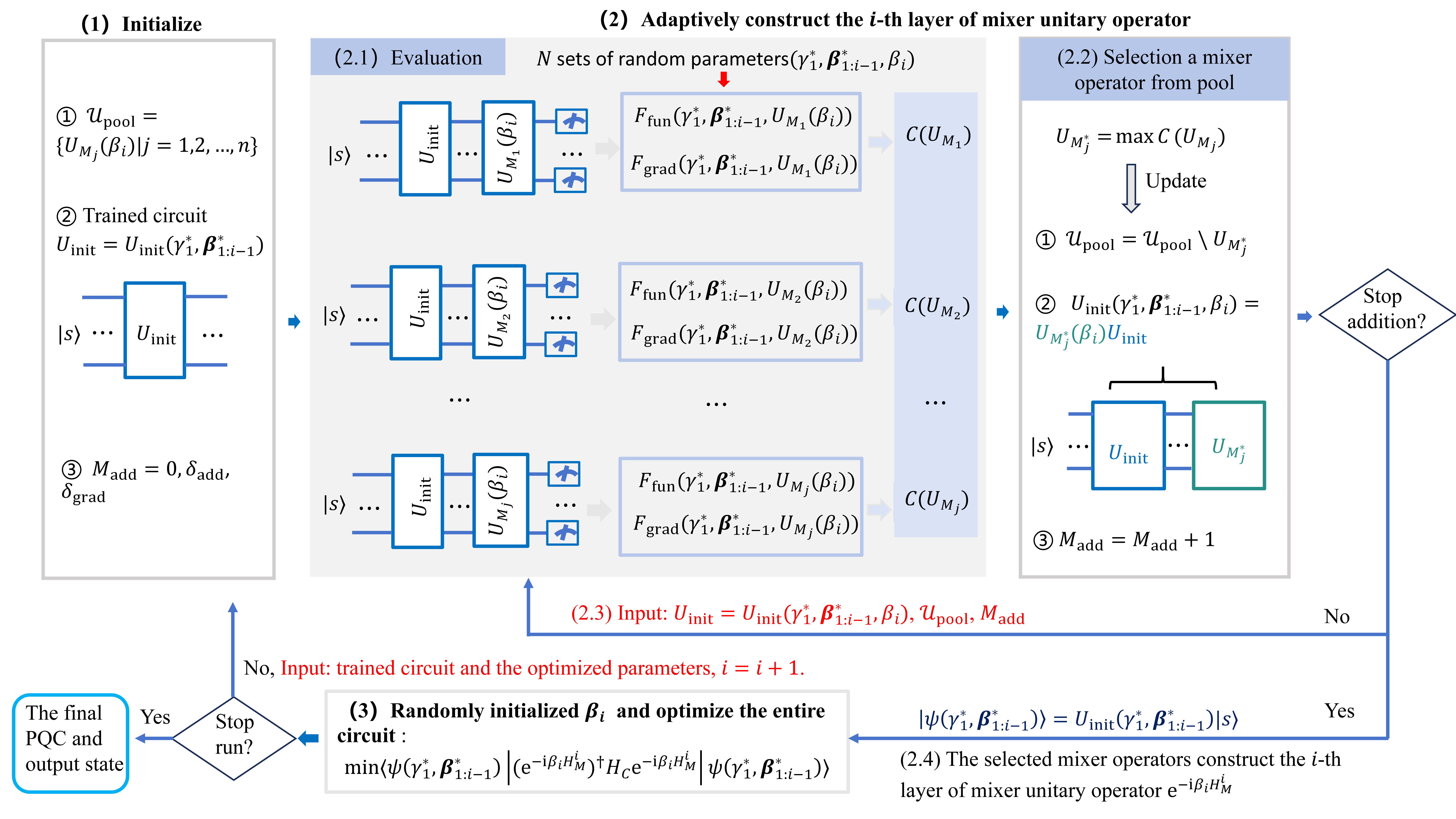}} % 替换为子图的文件名
	\end{minipage}
	\captionsetup{justification=raggedright}  % 左对齐
	\caption{\textcolor{black}{Schematic illustration of the adaptive circuit expansion process in AMA-QAOA+. The variable \( M_{\text{add}} \) is initialized to 0 to track the number of mixer operators added to the current layer of the mixer unitary operator. Two predefined thresholds, \( \delta_{\text{add}} \) and \( \delta_{\text{grad}} \), are given to control the number of added mixer operators. At each expansion, the evaluation function \( C(U_{M_j}) \) is computed for every candidate mixer operator \( U_{M_j} \) in the operator pool, where \( C(U_{M_j}) \) incorporates both the average initial expectation value and the average gradient. The mixer operator with the highest evaluation score is selected and appended to the current circuit $U_{\text{init}}$, after which the operator pool is updated and \( M_{\text{add}} \) is incremented by one. If the stopping conditions (i.e., $M_{\text{add}} \ge \delta_{\text{add}} $ or max($F_\text{grad}(U_{M_j}))<\delta_{\text{grad}}$ are not yet satisfied, the updated circuit $U_{\text{init}}$, operator pool, and \( M_{\text{add}} \) are passed into the next round of evaluation. This expansion process is repeated until the stop criterion is met, and these selected mixer operators commonly construct a new layer of the mixer unitary operator. Then, optimize the entire circuit, where AMA-QAOA+ randomly generates $\beta_{i}$, and reuses the optimized parameters $(\gamma_{1}^{*},\boldsymbol{\beta}_{1:i-1}^{*})$. If the stop condition in Equation~\eqref{eq:stopping_condition} is not met, re-initialize the relevant variables and continue constructing the next layer of the mixer unitary operator.}}
	% 整体图的标题
	\label{AMA_QAOA+_expansion} % 整体图的标签
\end{figure*}

\subsection{Optimize the entire circuit}
\textcolor{black}{After constructing the $i$-th layer of the mixer unitary operator, AMA-QAOA+ optimizes the newly obtained entire circuit. The initial parameters of the circuit are
	$(\gamma_{1}^{*},\beta_{1:i-1}^{*},\beta_{i})$, where $\beta_{i}$ is randomly initialized. After the optimization round, AMA-QAOA+ continues to construct the next mixer layer based on the newly optimized circuit $U_{\text{init}}(\gamma_{1}^{*},\boldsymbol{\beta}_{1:i-1}^{*})$ and parameters if the stopping condition in Equation~\eqref{eq:stopping_condition} is not met.} \textcolor{black}{To help readers better understand our work, the schematic illustration of the adaptive circuit expansion process in AMA-QAOA+ is given in \textbf{Figure~\ref{AMA_QAOA+_expansion}}. Besides, an outline of the AMA-QAOA+ procedures is shown in \textbf{Algorithm~\ref{AMA_steps}}.}

%Notably, considering the impact of different mixer arrangements on the circuit's optimization, AMA-QAOA+ selects mixers sequentially based on the evaluation function, rather than adding multiple mixers at once in a random order. This approach may ensure that the mixers are added in a systematic manner.

\medskip
All in all, though AMA-QAOA+ adaptively selects mixer operators from a predefined operator pool and iteratively grows the quantum circuit like the existing adaptive algorithms, there are some differences \textcolor{black}{as follows}.

\begin{itemize}
	
	\item \textbf{Evaluation difference.} The other adaptive algorithms use the gradient \textcolor{black}{of the expectation function concerning the newly introduced parameter under a set of random or fixed initial parameters} as the evaluation criterion. In contrast, AMA-QAOA+ uses the average gradient and average initial expectation values over multiple sets of parameters. The design of the evaluation function provides two main advantages in the selection process. First, it reduces the influence of randomness that could arise from using a single set of initial parameters by considering multiple sets of random initial parameters. Second, it considers both the optimization direction (via the gradient) and the quality of the solution (via the initial expectation function), which may lead to better results compared with methods relying on only one factor. \textcolor{black}{In addition, we investigate the impact of incorporating \( F_{\text{fun}} \) into the evaluation function on the performance of AMA-QAOA+ in Appendix~\ref{F_fun_effect}. Numerical results demonstrate that incorporating the average initial expectation value factor into the evaluation function enables the AMA-QAOA+ algorithm to achieve reduced resource consumption in terms of both CNOT gates and iterations, compared with implementation disregarding this critical factor.}
	
	\item \textbf{Optimization difference.} \textcolor{black}{The other adaptive algorithms optimize the overall circuit immediately after a single operator is added to the optimized circuit, where each selected operator is assigned an independent parameter. In contrast, AMA-QAOA+ optimizes the overall circuit after multiple operators are added to the optimized circuit, and these mixer operators share the same parameter \( \beta_i \).} \textcolor{black}{This training scheme is referred to as ``intermittent optimization”,} which may reduce the number of optimization rounds and the number of parameters to be optimized.
	
	\item \textbf{\textcolor{black}{Circuit structure difference.}} \textcolor{black}{Adaptive-QAOA \cite{Adaptive_qaoa, dynamic_adaptive_qaoa} follows the standard alternating structure of QAOA, while AMA-QAOA+ omits the target unitary operator after the first layer. The simulation results in Appendix~\ref{com} show that, compared with the implementation retaining this alternating structure, this structural simplification reduces the number of parameters to be optimized in AMA-QAOA+, thereby decreasing the dimensionality of the parameter space and facilitating faster convergence within each optimization round.}
	
\end{itemize}

\section{Numerical results} \label{simulation}
\textcolor{black}{
	This section presents a comprehensive comparison of algorithmic performance. We first describe the tested graph instances in subsection~\ref{dataset}, followed by a detailed explanation of the experimental settings in subsection~\ref{experiment settings}. The evaluation metrics used for comparison are introduced in subsection~\ref{comparison metrics}, and a thorough analysis of the numerical results is provided in subsection~\ref{analysis}.
}

\subsection{\textcolor{black}{Dataset}} \label{dataset}
We conducted numerical simulations on randomly generated ER graphs and 3-regular graphs, each with sizes \( n = 8, 10, 12, \textcolor{black}{14}\). \textcolor{black}{The ER random graph, denoted as $G(n, p_{0})$, is a graph with $n$ vertices, and $p_{0}$ is the probability that an edge exists between any two vertices, with each edge being chosen independently. In this paper, we chose $p_{0} = 0.5$, which ensures that the graphs generated are relatively dense and represent challenging instances for the MIS problem.} For each graph size, we generated 20 ER graphs and 20 3-regular graphs, resulting in a total of \textcolor{black}{160} test instances.

\subsection{\textcolor{black}{Experiment settings}}\label{experiment settings}

\textcolor{black}{All quantum algorithms begin with the initial state \( |0\rangle^{\otimes n} \), which is a feasible and easily prepared quantum state. For QAOA+ and PNU, the circuit layer depth is set to \( p = 3 \) and \( p = 5 \). In the case of PNU, the number of mixers per layer is fixed at \( N_\text{pm} = \lfloor n/2 \rfloor + 1 \), and the qubits acted upon by these mixers are randomly selected at each layer.}

\medskip
\textcolor{black}{For Adaptive-QAOA+ and AMA-QAOA+, a pre-trained parameterized quantum circuit is constructed by randomly selecting \( \lfloor n/2 \rfloor + 1 \) mixers prior to the adaptive construction process. The outer-loop termination criterion for both algorithms is defined as the absolute change in the expectation value being less than 0.1 across three consecutive optimization rounds (i.e., $\delta = 0.1$). Within each optimization round, convergence is declared when the absolute change in the expectation value is less than 0.01 over three successive iterations.}

\medskip
\textcolor{black}{The parameter $f_{1}$ balances the contribution of the expectation-based term ($F_{\text{fun}}$) and the gradient-based term ($F_{\text{grad}}$) in the evaluation function. Its core objective is to ensure that the algorithm effectively avoids the premature stagnation observed in purely gradient-based methods ($f_{1} \to 1$) within complex, flat energy landscapes. In our numerical exploration, we compute the average gradient and average initial expectation value using 10 sets of randomly initialized parameters, and we found $f_{1}=1/3$ was empirically found to be a good compromise, effectively compensating for the difference in scale between $F_{\text{fun}}$ and $F_{\text{grad}}$ in the current unnormalized setting. This balanced strategy significantly improves the algorithm's AR compared to purely gradient-based strategies, while simultaneously constructing more compact circuits, as evidenced by the numerical results in \textbf{Appendix~\ref{F_fun_effect}}.}

\medskip
\textcolor{black}{$\delta_{\text{add}}$ controls the number of partial mixers permitted to be added to a layer of the mixer unitary $U_M(\beta_{i})$. Its design goal is to manage circuit complexity and reduce the number of parameter optimization rounds. In our exploration, we found that as $\delta_{\text{add}}$ is set larger, the number of mixers allowed to be added to the circuit increases, which in turn leads to a significant increase in the CNOT gate count required by the final parameterized quantum circuit. This aligns with the intuitive expectation that larger $\delta_{\text{add}}$ leads to higher circuit complexity. AMA-QAOA+ employs a strategy of ``add multiple mixers, then optimize the circuit once." Compared to Adaptive-QAOA+'s strategy of ``add one mixer, then optimize the circuit once," appropriately increasing $\delta_{\text{add}}$ can save multiple rounds of parameter optimization without introducing excessive redundant CNOT gates, thus significantly boosting overall efficiency. }

\medskip
\textcolor{black}{$\delta_{\text{grad}}$ determines whether an existing mixer should be removed (pruned) based on its gradient contribution being close to zero (indicating redundancy). It is crucial for controlling circuit succinctness. In our exploration, we found that as $\delta_{\text{grad}}$ increased, the CNOT gate count required by the final converged circuit decreased, and the number of iterations also decreased. However, this comes at the cost of the algorithm converging to a lower-quality solution. This happens because setting $\delta_{\text{grad}}$ too high leads to the erroneous removal of mixers that have potential optimization value but currently weak gradient contributions. This premature pruning compromises the circuit's ability to perform fine-tuning, especially in later optimization stages, thereby sacrificing solution quality.}

\medskip
\textcolor{black}{The hyperparameters $\delta_{\text{add}}$ and $\delta_{\text{grad}}$ can control the number of partial mixers in a layer of the mixer unitary operator, fundamentally governing the trade-off between optimization efficiency and circuit complexity. In our final setting, we do not use a single fixed value. Instead, we implement a \textbf{stage-dynamic threshold strategy} based on the observation that the needs of the circuit change as it deepens.}
\begin{itemize}
	\item \textcolor{black}{\textbf{Early Stages (Looser Threshold):} In the earlier layers, we adopt looser thresholds (larger $\delta_{\text{add}}$ and $\delta_{\text{grad}}$) to allow more high-potential candidate mixers to be included. This broad exploration accelerates the initial convergence of the expectation function.}
	\item \textcolor{black}{\textbf{Later Stages (Tighter Threshold):} As the circuit deepens, we progressively tighten the thresholds (smaller $\delta_{\text{add}}$ and $\delta_{\text{grad}}$) to control the growth of the mixer more finely. This limits redundant gate additions and ensures the circuit remains compact.}
\end{itemize}

\textcolor{black}{This dynamic approach is essential to balance the efficiency requirement of \textbf{avoiding circuit overgrowth} (which saves CNOT gates) against the need to \textbf{reduce optimization rounds} (which accelerates convergence). The specific values used in our experiments for each stage are summarized in \textbf{}Table~\ref{tab:thresholds}. Our abundant numerical exploration results show that the current hyperparameter settings ensure the trade-off between the solution quality obtained by the AMA-QAOA+ algorithm and the consumption of CNOT gates and iteration counts.}

%\textcolor{black}{In AMA-QAOA+, we set \( f_1 = \frac{1}{3} \). During each evaluation, we compute the average gradient and average initial expectation value using 10 sets of randomly initialized parameters. To better manage the growth of the circuit while improving the quality of the solution, we dynamically adjust the threshold parameters \(\delta_{\text{grad}}\) and \(\delta_{\text{add}}\) based on the current number of mixer layers. In the earlier stages, we adopt looser thresholds to allow more candidate mixers to be included, accelerating the convergence of the expectation function. As the circuit deepens, we progressively tighten the thresholds to control the growth of the mixer more finely, thus limiting redundant gate additions and maintaining a compact circuit. The specific values used in our experiments are summarized in Table~\ref{tab:thresholds}.}

\begin{table}[h]
	\centering
	\renewcommand{\arraystretch}{1.13} % 增加行间距（默认是 1.0）
	\caption{\textcolor{black}{The threshold settings of $\delta_{\text{grad}}$ and $\delta_{\text{add}}$ in AMA-QAOA+.}}
	\label{tab:thresholds}
	\begin{tabular}{c@{\hspace{15pt}}c@{\hspace{15pt}}c}
		\hline
		\textbf{\textcolor{black}{The $i$-th Mixer Layers}} & \textcolor{black}{\(\delta_{\text{grad}}\)} & \textcolor{black}{\(\delta_{\text{add}}\)} \\
		\hline
		\textcolor{black}{2} & \textcolor{black}{$10^{-2}$} & \textcolor{black}{\(\lfloor n/2 \rfloor\)} \\
		
		\textcolor{black}{3} & \textcolor{black}{$10^{-1}$} & \textcolor{black}{\(\lfloor n/3 \rfloor\)} \\
		
		\textcolor{black}{\(\geq 4\)} & \textcolor{black}{$10^{-1}$} & \textcolor{black}{\(\lfloor n/4 \rfloor\)} \\
		\hline
	\end{tabular}
\end{table}

\medskip
The numerical simulations were conducted using the MindSpore Quantum 0.7.0 framework \cite{mindquantum}. We employ the Adam optimizer to train the QAOA parameters. Its ability to adapt learning rates for each parameter and incorporate momentum enables faster convergence and improved stability. The data that support the findings of this study are openly available in Github at \url{https://github.com/mindspore-lab/models/tree/master/research/arxiv_papers}.

\subsection{\textcolor{black}{Comparison Metrics}} \label{comparison metrics}
\textcolor{black}{ The optimal approximation ratio (OAR) reflects the best solution quality achieved by the algorithm among multiple optimization runs, and it is expressed as} 
\begin{equation}
	r_{\mathrm{opt}} = \max \{ r_{1}, r_{2}, \dots, r_{R} \},
\end{equation}
\textcolor{black}{where \( R \) is the total number of independent optimization runs. The average approximation ratio (AAR)}
\vspace{-0.5em}
\begin{equation}
	r_{\text{avg}} = \frac{1}{R} \sum_{t=1}^{R} r_{t},
	\vspace{-0.5em}
\end{equation}
\textcolor{black}{reflects the solution stability among multiple runs. Here, $r_{t}$ is the approximation ratio obtained in $t$-th optimization runs. A larger AAR indicates that the algorithm consistently produces high-quality solutions across different runs, and it has a greater stability.}

\medskip
\textcolor{black}{On a given type with $K$ graph instances, the mean OAR of an algorithm can be expressed as}
\vspace{-0.5em}
\begin{equation}
	\vspace{-0.5em}
	\overline{r}_{\mathrm{opt}} = \frac{1}{K} \sum_{k=1}^{K} r_{\mathrm{opt}}^{(k)},
	\vspace{-0.3em}
\end{equation}
\textcolor{black}{where $r_{\text{opt}}^{k}$ denotes the optimal approximation ratio achieved over multiple optimization runs on the $k$-th graph. The mean AAR of an algorithm can be expressed as }
\vspace{-0.5em}
\begin{equation}
	\overline{r}_{\mathrm{avg}} = \frac{1}{K} \sum_{k=1}^{K} r_{\mathrm{avg}}^{(k)}.
	\vspace{-0.5em}
\end{equation}
\textcolor{black}{Larger values of mean OAR indicate that the algorithm is more likely to consistently produce high-quality solutions across different instances. Moreover, a higher mean AAR also implies that the algorithm has greater solution stability across different instances and optimization runs.}

\medskip
\textcolor{black}{ To quantify the optimal solution quality and solution stability of an algorithm over a set of graph instances, we set $R = 100$ and analyze their performance based on the mean OAR and mean AAR. In addition, we also give the mean number of iterations required for parameter optimization per run on given graph types, as well as the mean number of CNOT gates required to implement the final constructed parameterized quantum circuit per run. These two metrics reflect the quantum resource consumption of the algorithm. A lower mean iteration counts imply faster convergence during parameter optimization, while smaller mean CNOT counts indicate that the resulting quantum circuit is more compact. The relevant definitions are as follows.}

\medskip
\textcolor{black}{
	On a given graph type with \( K \) graph instances, the \textit{mean number of iterations} is defined as
	\vspace{-0.5em}
	\begin{equation}
		\overline{T}_{\mathrm{iter}} = \frac{1}{K} \sum_{k=1}^{K} T_{\mathrm{iter}}^{(k)},
		\vspace{-0.5em}
	\end{equation}
}
where \( T_{\mathrm{iter}}^{(k)} \) denotes the average number of iterations consumed by the algorithm over $R$ optimization runs on the \( k \)-th graph instance. Specifically, 
\vspace{-0.5em}
\begin{equation}
	T_{\mathrm{iter}}^{(k)} = \frac{1}{R} \sum_{t=1}^{R} T_{\mathrm{iter}}^{(k,t)},
	\vspace{-0.5em}
\end{equation}
where \( T_{\mathrm{iter}}^{(k,t)} \) denotes the number of iterations consumed during the \( t \)-th optimization run of the algorithm on the \( k \)-th graph instance. Similarly, the \textit{mean number of CNOT gates} is given by
\vspace{-0.5em}
\begin{equation}
	\overline{T}_{\mathrm{CNOT}} = \frac{1}{K} \sum_{k=1}^{K} T_{\mathrm{CNOT}}^{(k)},
	\vspace{-0.5em}
\end{equation}
where \( T_{\mathrm{CNOT}}^{(k)} \) represents the average number of CNOT gates consumed by the algorithm over $R$ optimization runs on the \( k \)-th graph instance. Specifically, 
\vspace{-0.5em}
\begin{equation}
	T_{\mathrm{CNOT}}^{(k)} = \frac{1}{R} \sum_{t=1}^{R} T_{\mathrm{CNOT}}^{(k,t)},
	\vspace{-0.65em}
\end{equation}
where \( T_{\mathrm{CNOT}}^{(k,t)} \) denotes the number of CNOT gates consumed during the \( t \)-th optimization run of the algorithm on the \( k \)-th graph instance. In our calculation, the CNOT gate counts are calculated by decomposing each multi-qubit controlled $R_{x}$ gate into basic gates as outlined in Ref.~\cite{gate_decomposition_quantum}.

\subsection{\textcolor{black}{Experiments analysis}} \label{analysis}
\subsubsection{\textcolor{black}{Approximation ratio analysis on ER graphs}}
\begin{figure*}[htbp]
	\centering
	% 子图1
	\begin{minipage}[b]{0.458\textwidth} % 设置子图宽度为总宽度的45%
		\centering
		\subfloat[ Mean OAR, ER graphs.]{\label{OAR_ER} \includegraphics[width=\textwidth]{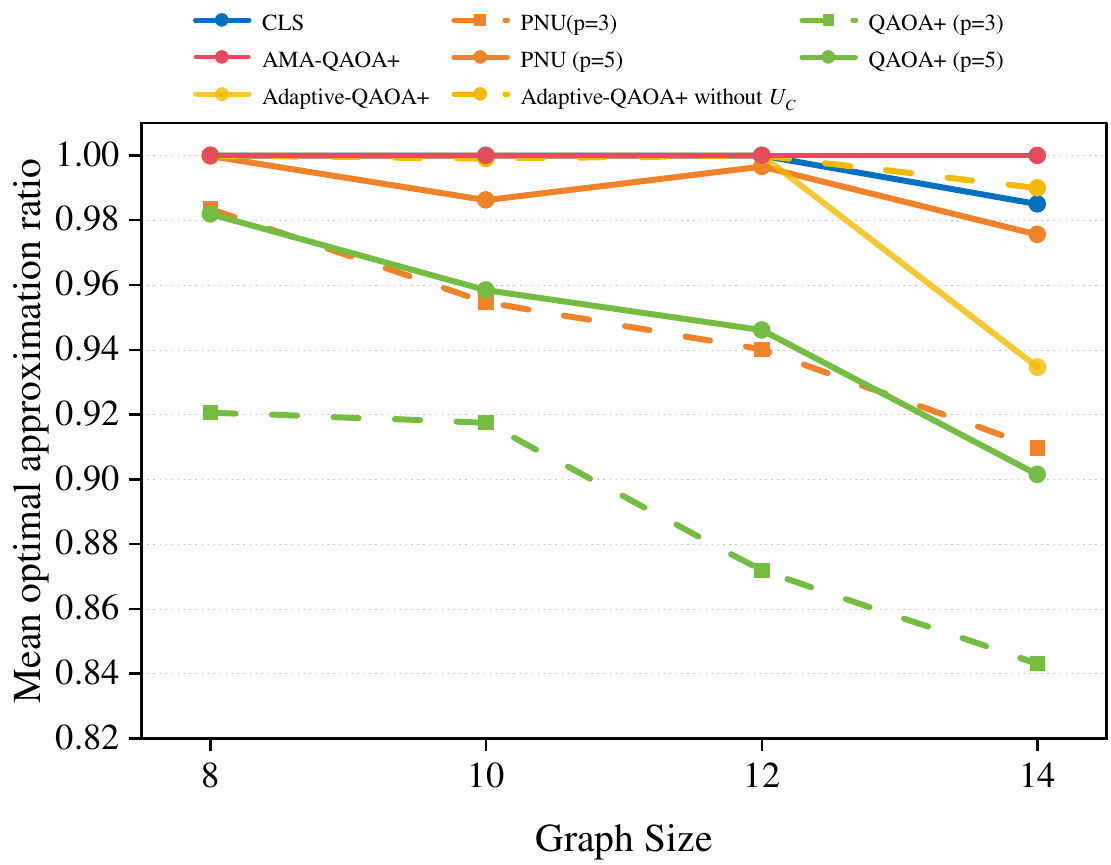}} % 替换为子图的文件名
	\end{minipage}
	\hspace{0.05\textwidth} % 调节间距
	% 子图2
	\begin{minipage}[b]{0.458\textwidth}
		\centering
		\subfloat[ Mean AAR, ER graphs.]{\label{AAR_ER} \includegraphics[width=\textwidth]{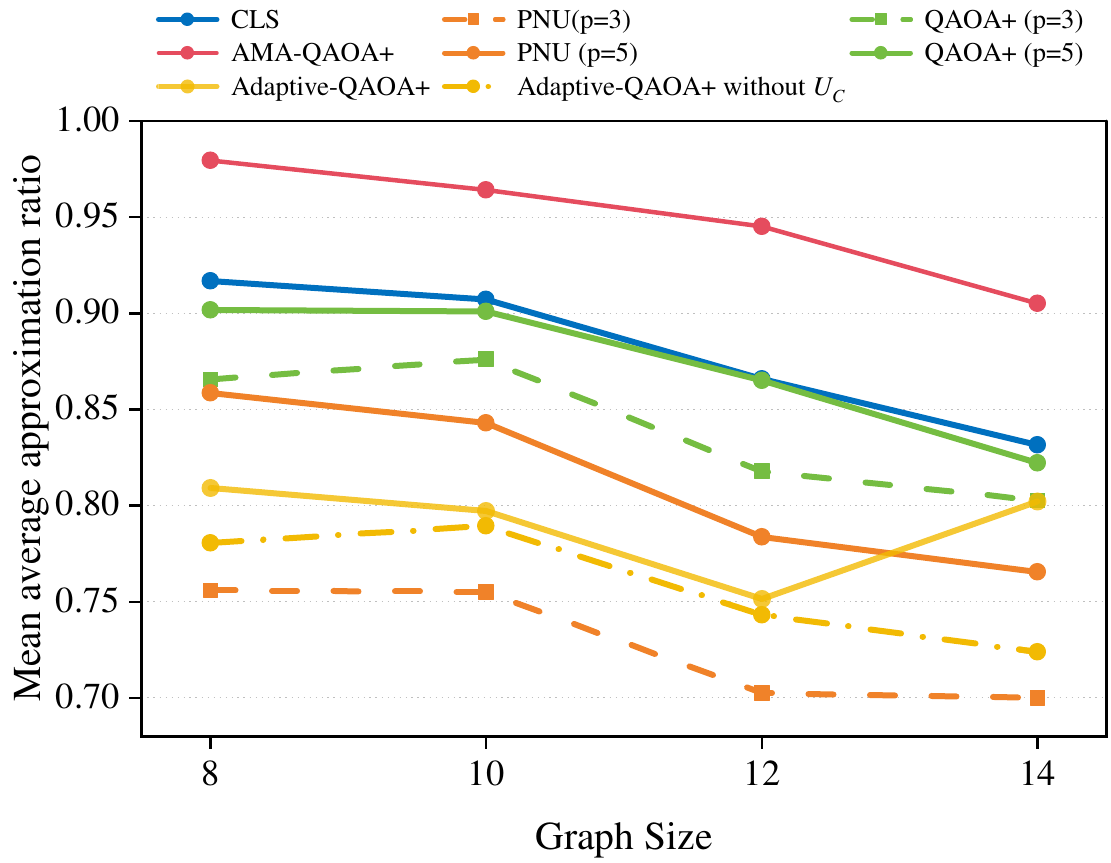}} % 替换为子图的文件名
	\end{minipage}
	\captionsetup{justification=raggedright, singlelinecheck=false}
	\setlength{\belowcaptionskip}{-0.45cm}   %调整图片标题与下文距离
	\caption{\textcolor{black}{The mean OAR and AAR obtained by various algorithms vary with graph size $n$ on ER graphs with an edge probability of 0.5. Here, ``Adaptive-QAOA+ without $U_{C}$'' is the variant of Adaptive-QAOA+, which is without the target unitary operator in $i(\ge 2)$-th layer of ansatz.}} % 整体图的标题
	\label{AR_ER} % 整体图的标签
\end{figure*}

\textbf{Figure~\ref{AR_ER}} presents the mean OAR and AAR obtained by different algorithms varying with the graph size, where each data point is obtained by averaging the data about 20 random graph instances. We observe that the PNU algorithm generally achieves a higher $\overline{r}_{\mathrm{opt}}$ than QAOA+ at the same layer depth and graph size. In addition, PNU attains the $\overline{r}_{\mathrm{opt}}$ close to 1 when the layer depth is set to \(p = 5\) and the graph size is \(n = 8,12\). These results suggest that some mixer allocation may be redundant in the original QAOA+, as comparable or even higher-quality solutions can be achieved with fewer mixers. However, in terms of the $\overline{r}_{\mathrm{avg}}$ under the same layer depth and the same number of optimization runs, QAOA+ demonstrates better stability. This is because PNU lacks an effective mixer selection strategy, and it randomly selects mixers in each optimization run. As a result, its performance fluctuates considerably across runs and exhibits limited stability. In summary, while PNU can reduce mixer consumption through random mixer selection, it suffers from solution instability. In contrast, QAOA+ applies mixers uniformly to all qubits, ensuring structural consistency but introducing redundancy in mixer usage. These observations highlight the requirement of designing adaptive mixer selection strategies to balance resource efficiency and solution quality.

\medskip
\textcolor{black}{As the graph size increases, the mean approximation ratios achieved by all algorithms generally exhibit a downward trend. This is primarily due to the increased structural complexity of larger graphs, which expands the solution space and makes the optimization process more susceptible to local minima. Despite this overall decline, AMA-QAOA+ consistently achieves higher mean optimal and mean average approximation ratios than all other algorithms, demonstrating both strong solution quality and stability. In contrast, although Adaptive-QAOA+ occasionally achieves a higher $\overline{r}_{\mathrm{opt}}$ than QAOA+ and PNU, its $\overline{r}_{\mathrm{avg}}$ is often lower than that of QAOA+ with $p = 3,5$ and PNU with $p = 5$. The low $\overline{r}_{\mathrm{avg}}$ suggests that Adaptive-QAOA+ exhibits high variance across multiple optimization runs or graph instances, leading to reduced solution stability. This instability may be attributed to its single-shot evaluation strategy, which may result in overlooking critical mixers or prematurely including less effective ones, and the evolution of the quantum state tends to be misdirected to low-quality solutions.}

\begin{figure*}[htbp]
	\centering
	% 子图1
	\begin{minipage}[b]{0.458\textwidth} % 设置子图宽度为总宽度的45%
		\centering
		\subfloat[ Mean CNOT gates, ER graphs.]{\label{CNOT_ER} \includegraphics[width=\textwidth]{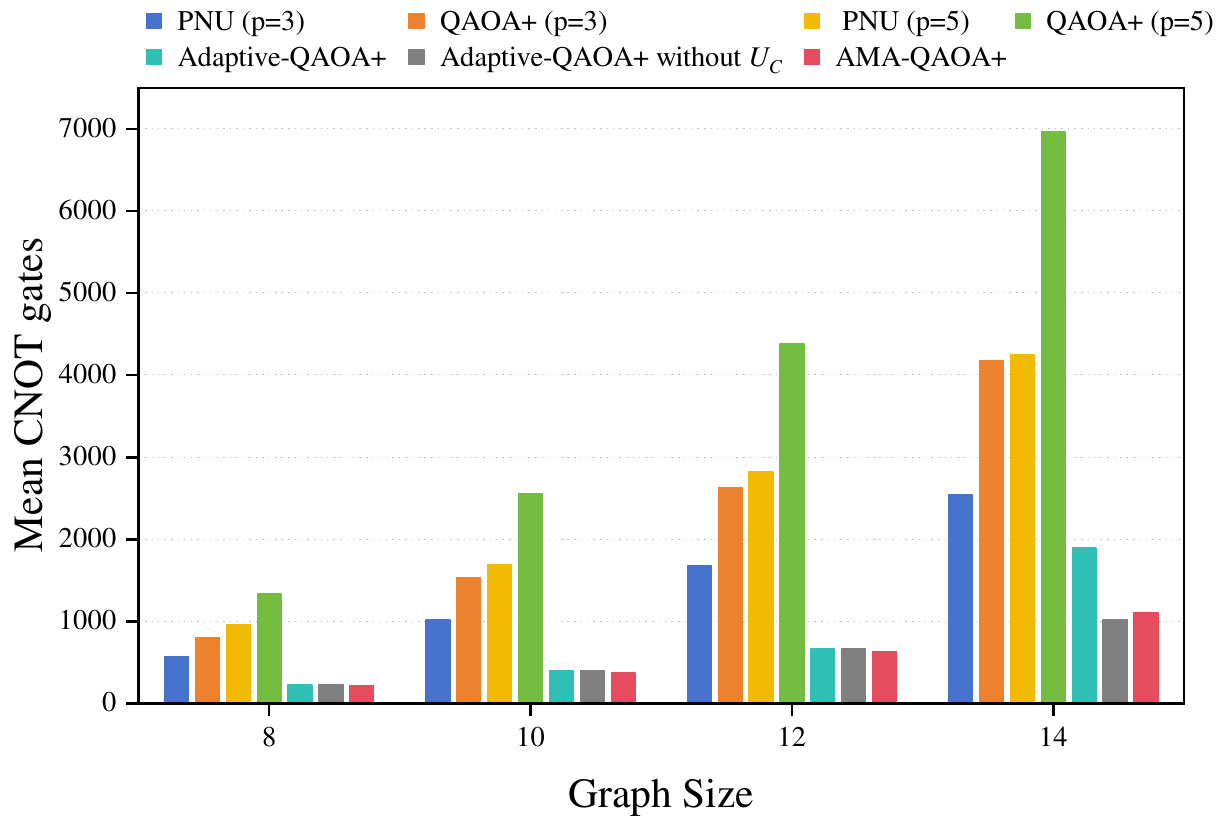}} % 替换为子图的文件名
	\end{minipage}
	\hspace{0.05\textwidth} % 调节间距
	% 子图2
	\begin{minipage}[b]{0.458\textwidth}
		\centering
		\subfloat[ Mean iterations, ER graphs.]{\label{ITR_ER} \includegraphics[width=\textwidth]{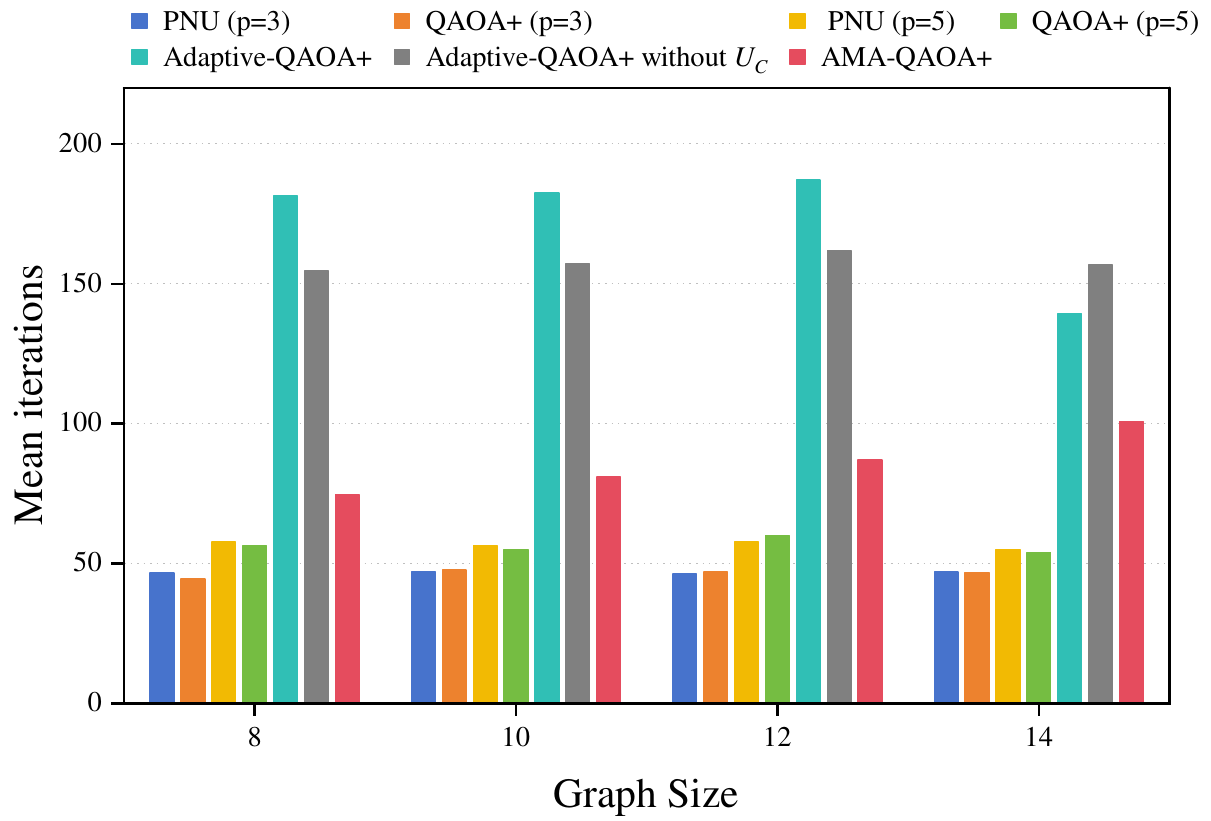}} % 替换为子图的文件名
	\end{minipage}
	
	\captionsetup{justification=raggedright, singlelinecheck=false}
	\setlength{\belowcaptionskip}{-0.45cm}   %调整图片标题与下文距离
	\caption{\textcolor{black}{The mean number of CNOT gates and iterations required by different algorithms per run varies with the graph size $n$ on ER graphs with an edge probability of 0.5.  To facilitate a more direct comparison of how the evaluation function and optimization method influence resource consumption, this figure also includes a variant of Adaptive-QAOA+ that excludes the target unitary operator in the \( i \)-th layer for \( i \geq 2 \).}} % 整体图的标题
	\label{resource_ER} % 整体图的标签
\end{figure*}

\medskip
\textcolor{black}{ Beyond comparisons among quantum algorithms, we further introduce the classical local search (CLS) algorithm \cite{local_search_MIS} as a baseline to evaluate solution quality and stability under the same number of optimization runs. Numerical results show that CLS outperforms QAOA+, PNU, and Adaptive-QAOA+ in terms of $\overline{r}_{\mathrm{opt}}$ and $\overline{r}_{\mathrm{avg}}$. In contrast, AMA-QAOA+ consistently achieves higher mean optimal and average approximation ratios than CLS across all tested graph types, confirming its superiority over it.
}
%Beyond comparisons among quantum algorithms, it is also important to examine whether quantum methods can outperform classical heuristics under same runs. To this end, we additionally include the CLS algorithm and a greedy algorithm for the MIS problem as baselines for evaluating solution quality and stability. The greedy algorithm, referred to as greedy-MIS, consistently achieves both an optimal and average approximation ratio of 1 across all tested graph instances. As its performance remains constant and dominates the y-axis range, we omit its curves in the figure for visual clarity. The results demonstrate that while CLS outperforms QAOA+\((p=3)\) and PNU\((p=3)\) in terms of average approximation ratio, AMA-QAOA+ outperforms CLS in both solution quality and stability across all tested instances, confirming its superiority over it.

\subsubsection{\textcolor{black}{Resource consumption  analysis of ER graphs}}
%\textcolor{black}{In addition to assessing solution quality, evaluating the computational resources per run required by each algorithm is important, which can help evaluate their scalability and practical feasibility on near-term quantum hardware. In this subsection, we examine the mean number of CNOT gates per run required to implement the final parameterized quantum circuit, as well as the mean number of iterations per run required for parameter optimization. }

\medskip
\textbf{Figure~\ref{CNOT_ER}} shows that all algorithms exhibit a quick growth in the number of CNOT gates required per run as the graph size increases. This trend is particularly pronounced for QAOA+, which applies mixers to all qubits. As a result, a significantly higher number of CNOT gates is needed when decomposing each mixer into elementary gates \textcolor{black}{compared with} other algorithms. In contrast, Adaptive-QAOA+ and AMA-QAOA+ restrict mixers to selected qubits, substantially lowering CNOT gate counts. Furthermore, AMA-QAOA+ \textcolor{black}{usually} consumes fewer CNOT gates than both Adaptive-QAOA+ and its variant across all graph sizes, with the gap widening as the graph size increases. This efficiency mainly stems from the mixer selection strategy of AMA-QAOA+, which evaluates candidate mixers based on both the gradient and the initial expectation value, averaged over multiple sets of random initial parameters. This evaluation method leads to more effective mixer selection, resulting in a more compact and resource-efficient circuit.

\medskip
\textbf{Figure~\ref{ITR_ER}} presents the mean number of iterations required for parameter optimization per run. We observe that PNU and QAOA+ consistently require fewer iterations per run than the adaptive algorithms at the same graph size. This is because adaptive algorithms typically involve multiple rounds of optimization within each run, \textcolor{black}{while QAOA+ and PNU only include an optimization round per run.} The multiple optimization rounds inevitably introduce additional iteration overheads. But for AMA-QAOA+, this resource sacrifice is exchanged for higher solution quality. While the number of iterations is significant, the number of CNOT gates required per run is more directly related to the problem size that can be supported by resource-constrained quantum devices. Excessive demands in CNOT gates may limit the capacity of the hardware to solve specific problems. Therefore, it can be a justified trade-off to sacrifice some iterations to reduce CNOT gate consumption. This approach allows the algorithm to effectively tackle issues within the limited available resources. Among all adaptive algorithms, AMA-QAOA+ exhibits fewer iterations than Adaptive-QAOA+ and its variant. This advantage is primarily from its optimization strategy. Specifically, AMA-QAOA+ postpones optimization until multiple mixer operators have been added, rather than optimizing after the insertion of each mixer operator. This scheme is referred to as ``intermittent optimization''.  Compared with the “single-add-single-optimize” strategy employed by Adaptive-QAOA+ and its variant, this approach reduces the frequency of intermediate parameter updates, thereby substantially lowering the total iteration counts.

\medskip
\textcolor{black}{In summary, the results on ER graphs with an edge probability of 0.5 show that AMA-QAOA+ achieves a desirable balance between solution quality and resource efficiency. It adaptively selects a subset of high-impact mixers based on averaged gradient values and initial expectation function values, significantly reducing CNOT gate consumption and improving the solution quality. Moreover, AMA-QAOA+ also adopts an intermittent optimization strategy, which reduces the number of iterations per run, thereby improving the efficiency of classical optimization. In conclusion, compared with Adaptive-QAOA+ and other variants, AMA-QAOA+ shows higher efficiency in mixer selection.
}
%These features make AMA-QAOA+ a scalable and practical candidate for implementation on near-term quantum devices with limited available resources.

\subsubsection{\textcolor{black}{Performance analysis on 3-regular graphs: Approximation ratio and resource consumption}}

\begin{figure*}[htbp]
	\centering
	% 子图1
	\begin{minipage}[b]{0.458\textwidth} % 设置子图宽度为总宽度的45%
		\centering
		\subfloat[ Mean OAR, 3-regular graphs.]{\label{OAR_regular} \includegraphics[width=\textwidth]{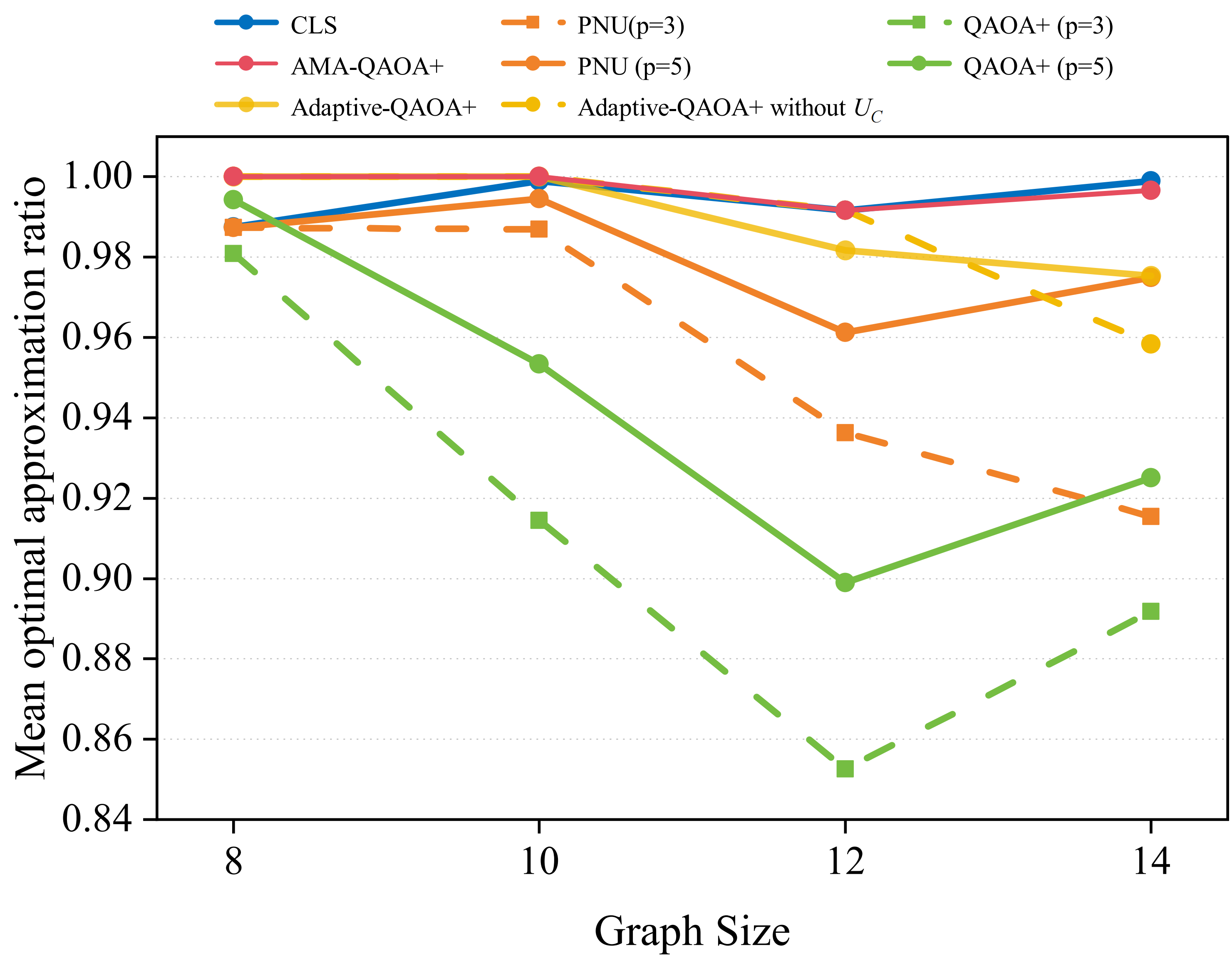}} % 替换为子图的文件名
	\end{minipage}
	\hspace{0.05\textwidth} % 调节间距
	% 子图2
	\begin{minipage}[b]{0.458\textwidth}
		\centering
		\subfloat[ Mean AAR, 3-regular graphs.]{\label{AAR_regular} \includegraphics[width=\textwidth]{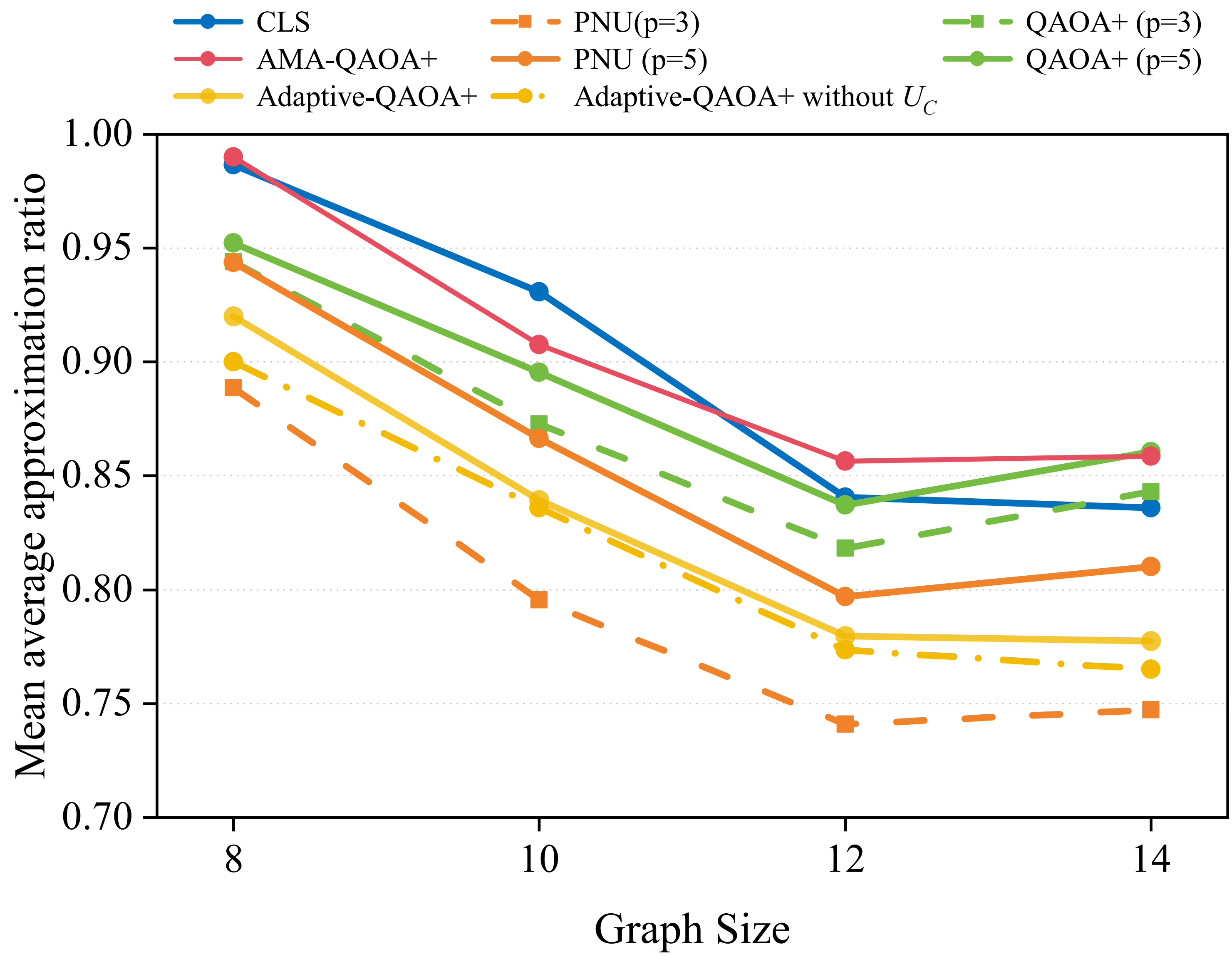}} % 替换为子图的文件名
	\end{minipage}
	\captionsetup{justification=raggedright, singlelinecheck=false}
	\setlength{\belowcaptionskip}{-0.45cm}   %调整图片标题与下文距离
	\caption{\textcolor{black}{The mean OAR and AAR obtained by various algorithms vary with graph size $n$ on 3-regular graphs. Here, ``Adaptive-QAOA+ without $U_{C}$'' is the variant of Adaptive-QAOA+, which is without the target unitary operator in $i(\ge 2)$-th layer.}} % 整体图的标题
	\label{AR_regular} % 整体图的标签
\end{figure*}

\begin{figure*}[htbp]
	\centering
	% 子图1
	\begin{minipage}[b]{0.458\textwidth} % 设置子图宽度为总宽度的45%
		\centering
		\subfloat[ Mean CNOT gates, 3-regular graphs.]{\label{CNOT_regular} \includegraphics[width=\textwidth]{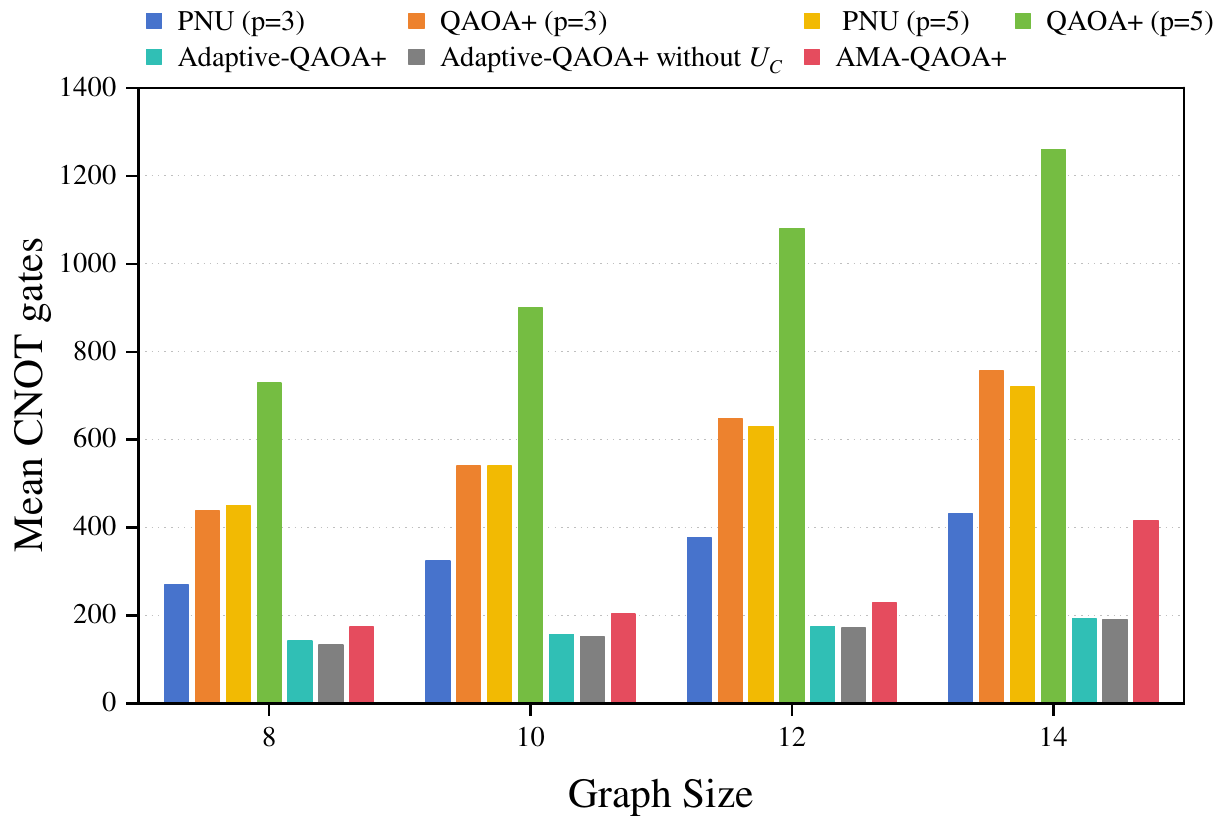}} % 替换为子图的文件名
	\end{minipage}
	\hspace{0.05\textwidth} % 调节间距
	% 子图2
	\begin{minipage}[b]{0.458\textwidth}
		\centering
		\subfloat[ Mean iterations, 3-regular graphs.]{\label{ITR_regular} \includegraphics[width=\textwidth]{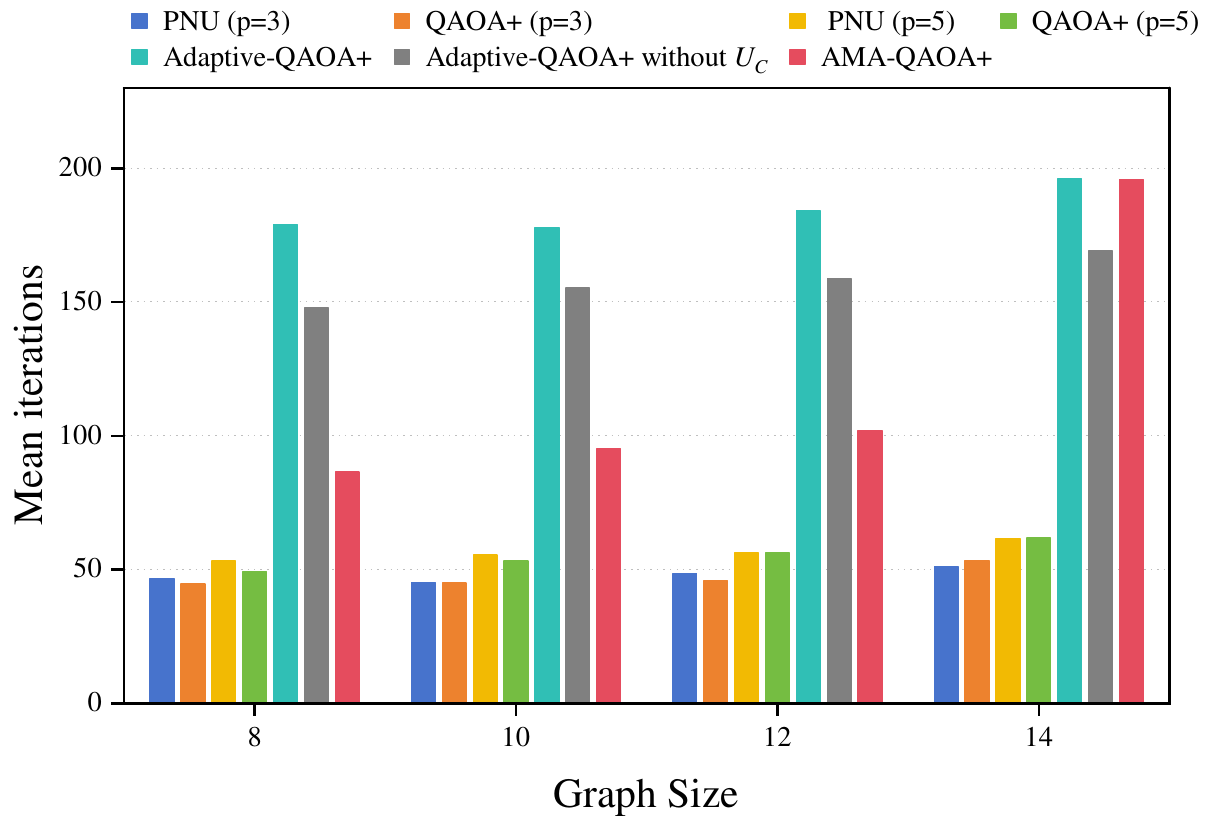}} % 替换为子图的文件名
	\end{minipage}
	\captionsetup{justification=raggedright, singlelinecheck=false}
	\setlength{\belowcaptionskip}{-0.45cm}   %调整图片标题与下文距离
	\caption{\textcolor{black}{The mean number of CNOT gates and iterations required by different algorithms per run varies with the graph size $n$ on 3-regular graphs.  To facilitate a more direct comparison of how the evaluation function and optimization method influence resource consumption, this figure also includes a variant of Adaptive-QAOA+ that excludes the target unitary operator in the \( i \)-th layer for \( i \geq 2 \).}
	} % 整体图的标题
	\label{resource_regular} % 整体图的标签
\end{figure*}

\textbf{Figure~\ref{AR_regular}} and \textbf{Figure~\ref{resource_regular}} respectively show the mean approximation ratios and mean resource usage (CNOT gates and iterations for optimization per run) for different algorithms on 3-regular graphs of varying sizes. The results reveal three key observations. First, the performance on 3-regular graphs further provides evidence of the redundant mixer allocation problem in the original QAOA+. Second, AMA-QAOA+ can still achieve higher-quality and more stable solutions than the compared quantum algorithms on 3-regular graphs. Third, unlike the ER graph results, AMA-QAOA+ does not consistently outperform Adaptive-QAOA+ and the variant of Adaptive-QAOA+ in terms of resource efficiency on 3-regular graphs. The detailed analysis is as follows.

\medskip
\textcolor{black}{Under the same layer depth (\(p=5\)) and the number of optimization runs, PNU achieves the $\overline{r}_{\mathrm{opt}}$ close to 1 in graph sizes of 8 and 10, outperforming QAOA+. This indicates that applying mixers to only a subset of qubits, as done in PNU, is sufficient to achieve quasi-optimal solutions. This phenomenon implies that the full-mixer allocation strategy in QAOA+ has unnecessary mixer overheads. Such excessive mixer usage increases circuit complexity, which not only raises the implementation cost but also hampers the effectiveness of parameter optimization, ultimately degrading the overall performance. This observation is consistent with the performance patterns previously observed on ER graphs of an edge probability of 0.5.}

\begin{table*}[htbp]
	\centering
	\renewcommand{\arraystretch}{1.15} % 增加行间距（默认是 1.0）
	\caption{\textcolor{black}{Relative performance changes (\%) of AMA-QAOA+ \textcolor{black}{compared with} other algorithms on ER graphs.}}
	\label{tab:er_resource_comparison}
	{\color{black}
		\begin{tabular}{lcccc}
			\hline
			\textbf{Algorithm} & $\overline{r}_{\mathrm{opt}} \uparrow$ & $\overline{r}_{\mathrm{avg}} \uparrow$ & $\overline{T}_{\mathrm{CNOT}} \downarrow$ & $\overline{T}_{\mathrm{iter}}$ \\
			
			\hline
			PNU (p=3)        &+5.30       & +20.97 & $-61.07$   & +83.80 \\
			QAOA+ (p=3)      &+11.16      & +9.77  & $-74.51$   & +84.41 \\
			PNU (p=5)        &+10.40      & +12.54 & $-76.71$   & +51.62 \\
			QAOA+ (p=5)      &+5.30       & +6.56  & $-84.70$   & +52.81 \\
			Adaptive-QAOA+   &+1.63       & +14.81 & $-14.61$   & $-48.90$ $\downarrow$\\
			Adaptive-QAOA+ (variant)&+0.28 & +17.88 & $-2.04$   & $-45.54$ $\downarrow$\\
			\hline
		\end{tabular}
	}
\end{table*}

\begin{table*}[htbp]
	\centering
	\renewcommand{\arraystretch}{1.15} % 增加行间距（默认是 1.0）
	\caption{\textcolor{black}{Relative performance changes (\%) of AMA-QAOA+ \textcolor{black}{compared with} other algorithms on 3-regular graphs.}}
	\label{tab:regular_resource_comparison}
	{\color{black}
		\begin{tabular}{lcccc}
			\hline
			\textbf{Algorithm} & $\overline{r}_{\mathrm{opt}} \uparrow$ & $\overline{r}_{\mathrm{avg}} \uparrow$ & $\overline{T}_{\mathrm{CNOT}}$ & $\overline{T}_{\mathrm{iter}}$ \\
			
			\hline
			PNU (p=3)        &+4.05      & +11.01 & $-28.95$ $\downarrow$ & +146.95 \\
			QAOA+ (p=3)      &+8.71      & +3.37  & $-58.03$ $\downarrow$ & +148.40 \\
			PNU (p=5)      &+1.76       & +4.89  & $-57.37$ $\downarrow$ & +108.37 \\
			QAOA+ (p=5)    &+5.41        & +1.69  & $-74.82$ $\downarrow$ & +113.22 \\
			Adaptive-QAOA+  &+0.78      & +7.41  & $+49.42$     & $-35.74$ $\downarrow$ \\
			Adaptive-QAOA+ (variant) &+0.96 & +8.45  & $+53.78$     & $-25.02$ $\downarrow$ \\
			\hline
		\end{tabular}
	}
\end{table*}

\medskip
\textcolor{black}{Moreover, AMA-QAOA+ consistently achieves the highest mean optimal and average approximation ratios among all quantum algorithms across all tested graph sizes, demonstrating its scalability and stability on 3-regular graphs. In contrast, Adaptive-QAOA+ and its variant can occasionally outperform QAOA+ and PNU in terms of $\overline{r}_{\mathrm{opt}}$. However, its $\overline{r}_{\mathrm{avg}}$ is generally lower than both QAOA+ and PNU (\(p=5\)) and tends to deteriorate further as the graph size increases. This instability stems primarily from its mixer selection strategy, which relies on the gradient evaluation computed from a single set of randomly initialized parameters. On regular graphs, the structural homogeneity among vertices makes the gradient landscape less distinctive. As a result, gradients computed from a single set of random initial parameters may fail to capture meaningful differences among candidate mixers, leading the algorithm to select suboptimal ones. These suboptimal mixers may bias the circuit construction process toward ineffective directions in the solution space, thereby increasing the likelihood of convergence to low-quality local minima. Once the algorithm becomes trapped in such local minima, the expectation function tends to improve only marginally during subsequent optimization rounds, often triggering termination conditions \textcolor{black}{given in Equation~\eqref{eq:stopping_condition}}.}

%While this premature termination may reduce the total number of iterations and CNOT gates, it ultimately limits the solution quality and undermines the stability of the Adaptive-QAOA+ algorithm.

\medskip
\textcolor{black}{ In terms of resource consumption, both Adaptive-QAOA+ and AMA-QAOA+ use significantly fewer CNOT gates per run than PNU and QAOA+ at the same graph size. This reduction stems from their selective mixer allocation strategy, which circumvents the expensive application of mixers to all qubits. A more detailed comparison reveals that Adaptive-QAOA+ and its variant consume fewer CNOT gates per run than AMA-QAOA+ on 3-regular graphs—a trend opposite to that observed on ER graphs of an edge probability of 0.5. This behavior results from premature saturation of the optimization process in Adaptive-QAOA+ and its variant. Owing to the structural homogeneity of 3-regular graphs and the reliance on gradient information from a single random initialization, Adaptive-QAOA+ tends to select ineffective mixers that fail to drive meaningful quantum state evolution. As a result, the expectation function quickly stagnates, and further circuit growth is suppressed. The optimization halts early, leading to reduced CNOT gate usage. However, this comes at the cost of lower solution quality and less stable performance, as reflected in the decreasing $\overline{r}_{\mathrm{avg}}$ with increasing graph size.}

\medskip
\textcolor{black}{In summary, while Adaptive-QAOA+ and its variant consume fewer CNOT gates due to early saturation, the results show that this early termination ultimately compromises solution quality and leads to unstable performance. In contrast, AMA-QAOA+ employs a more robust mixer evaluation function, and the results demonstrate that this evaluation can facilitate the reliable selection of high-impact mixers and mitigate the risk of premature convergence to low-quality local minima compared with a single-shot evaluation strategy. Although AMA-QAOA+ incurs a higher CNOT gate overhead per run \textcolor{black}{compared with} Adaptive-QAOA+ on 3-regular graphs, this additional cost is effectively leveraged to steer the quantum state toward more optimal regions of the solution space. Combined with its consistently superior mean OAR and AAR on 3-regular graphs, AMA-QAOA+ achieves a more favorable trade-off between solution quality and resource efficiency than all competing methods. These advantages underscore the scalability and stability of AMA-QAOA+.
}

\medskip
%ER图上，AMA-QAOA+相较于PNU（p=5）平均近似比提升了12.54%，相较于QAOA+（p=5）提升了6.56%，相较于Adaptive-QAOA+及其变体分别提升了14.81%以及17.88%。 CNOT门消耗方面，AMA-QAOA+相较于PNU（p=5）降低了76.71%，相较于QAOA+（p=5）降低了84.70%，相较于Adaptive-qaoa+及其变体分别降低了14.61%以及2.04%。迭代步数消耗方面，AMA-QAOA+相较于PNU（p=5）增加了51.62%，相较于QAOA+（p=5）增加了52.81%，相较于Adaptive-qaoa+及其变体分别降低了48.90%以及45.54%。

%3正则图上，AMA-QAOA+相较于PNU（p=5）平均近似比提升了4.89%，相较于QAOA+（p=5）提升了1.69%，相较于Adaptive-QAOA+及其变体分别提升了7.41%以及8.45%。 CNOT门消耗方面，AMA-QAOA+相较于PNU（p=5）降低了57.37%，相较于QAOA+（p=5）降低了74.82%。迭代步数消耗方面，AMA-QAOA+相较于PNU（p=5）增加了108.37%，相较于QAOA+（p=5）增加了113.22%，相较于Adaptive-qaoa+及其变体分别降低了35.74%以及25.02%。
\textcolor{black}{To provide a direct performance comparison of AMA-QAOA+ with other algorithms on the tested graph instances, \textbf{Tables~\ref{tab:er_resource_comparison}} and \textbf{~\ref{tab:regular_resource_comparison}} report the percentage improvements in $\overline{r}_{\mathrm{opt}}$, $\overline{r}_{\mathrm{avg}}$  
	and the changes in $\overline{T}_{\mathrm{CNOT}}$ and $\overline{T}_{\mathrm{iter}}$. The comparisons are based on the results from 160 tested instances. Positive values in the $\overline{r}_{\mathrm{opt}}$ and $\overline{r}_{\mathrm{avg}}$ columns indicate that AMA-QAOA+ achieves higher solution quality and better solution stability than the compared algorithms. The negative values in the $\overline{T}_{\mathrm{CNOT}}$ and $\overline{T}_{\mathrm{iter}}$ columns indicate the percentage of resource consumption reduced by AMA-QAOA+ relative to the compared algorithms, demonstrating improved implementation efficiency. Positive values in these resource columns indicate increased resource usage, which is often a trade-off for achieving higher solution quality or more stable optimization behavior. Overall, the results highlight the ability of AMA-QAOA+ to achieve superior solution quality and stability, often with reduced quantum resource requirements, thus demonstrating a favorable balance between performance and resource efficiency.
}

\section{Conclusion and Outlook}\label{conclusion}
To address the substantial gate overheads of QAOA+ in solving constrained combinatorial optimization problems, we proposed the AMA-QAOA+ algorithm. AMA-QAOA+ adaptively constructs a feasible solution space by selectively applying mixers to a subset of qubits in each layer of the mixer unitary operator, and then searches for the quasi-optimal solution within this constrained space. We evaluated the performance of AMA-QAOA+ on the MIS problem and compared it with QAOA+, PNU, Adaptive-QAOA+, \textcolor{black}{ and the variant of Adaptive-QAOA+. Extensive numerical simulations demonstrate that, under the same number of optimization runs, AMA-QAOA+ consistently achieves superior performance.
	Specifically, AMA-QAOA+ obtained higher mean OARs and AARs, indicating higher solution quality and greater stability, along with significantly reduced CNOT gates. These results highlight the feasibility and efficiency of AMA-QAOA+ in solving CCOPs.}

\medskip
The AMA-QAOA+ algorithm is built upon two core mechanisms, i.e., an evaluation-based mixer selection strategy and an intermittent optimization scheme. These mechanisms are problem-agnostic and can be extended to other CCOPs that are solvable within the QAOA+ framework. Adapting AMA-QAOA+ to new CCOPs primarily requires redesigning the target Hamiltonian and the mixer operator pool to accurately encode the new objective and associated constraints. For many representative CCOPs, such as the minimum set cover, maximum set packing, and minimum vertex cover, previous studies (as shown by Hadfield et al.) have designed mixer unitaries capable of generating a feasible solution space. These mixers often involve multi-qubit controlled gates, and applying them uniformly across all qubits can result in substantial gate overhead. In such scenarios, AMA-QAOA+ offers a promising approach for constructing more compact and resource-efficient quantum circuits. In summary, while this paper emphasizes the efficiency of AMA-QAOA+ in addressing the MIS problem, the algorithm itself is versatile and can be applied to various CCOPs. Future research should investigate its applicability to other CCOPs and assess its scalability across larger instances and different graph topologies, which would enhance its utility in a variety of optimization contexts.

\medskip
\textcolor{black}{Currently, the $F_{\text{fun}}$ and $F_{\text{grad}}$ terms in the evaluation function are utilized in their unnormalized scale, and they may not be on the same scale, which could bias the contribution of each term. It may be more appropriate to normalize them. For example, by rescaling each over $N$ random sets using minima and maxima of $F_\text{init}$ and partial derivative, so that both terms share a comparable range. Normalizing $F_{\text{fun}}$ and $F_{\text{grad}}$ to a comparable range is theoretically more rigorous for creating a robust, bias-free dual-criteria evaluation function. Future work should analyze the potential issues arising from this unnormalized scale and investigate how the hyperparameter $f_{1}$ should be set based on a robust normalization scheme to achieve a consistently optimal trade-off between solution quality and resource consumption across diverse problem instances. Furthermore, the current evaluation function is designed to be problem-agnostic. However, for graph structures with special topological properties, such as regular graphs or highly sparse/dense graphs, designing a new, topology-aware evaluation function could potentially prioritize the most influential mixers specific to that topology. This tailored approach could significantly enhance the efficiency and performance of AMA-QAOA+ on specialized problem sets.} Additionally, future efforts could focus on improving the optimization and initialization strategies for AMA-QAOA+. For instance, integrating deep learning techniques to predict high-quality initial parameters may speed up convergence and improve optimization efficiency, particularly in high-dimensional problem spaces. Furthermore, it is crucial to evaluate its performance on real quantum hardware to understand its behavior in realistic conditions. Such experiments would provide valuable insights into its practical advantages and limitations and inform further refinements of the algorithm.

%\textcolor{black}{Furthermore, although the current design of AMA-QAOA+ selects high-impact mixers based on averaged gradients and initial expectation values, recent works suggest that additional criteria—such as partial CNOT cancellation, hardware connectivity, and circuit depth contribution—can further guide mixer selection and ordering. For instance, arXiv:2209.10562 introduces structure-aware ansatz construction techniques aimed at minimizing circuit depth. Integrating such resource-aware criteria into the evaluation process could help AMA-QAOA+ better balance performance improvements against hardware limitations. In the future, we plan to investigate multi-objective evaluation strategies that combine performance metrics with gate cost and circuit depth. Such strategies may further enhance the scalability and practicality of AMA-QAOA+ on near-term quantum hardware.}

%The results not only verify the superior performance of AMA-QAOA+ but also demonstrate its adaptability and robustness in varying AR scenarios. These findings underscore its potential as an efficient algorithm, particularly in resource-constrained environments, such as quantum computing, where AMA-QAOA+ offers substantial practical value.

%%%%%%%%%%%%%%%%%%%%%%%%%%%%%%%%%%%%%%%%%%%%%%%%%%%%%%%
%%% Acknowledgements. 致谢
%%%%%%%%%%%%%%%%%%%%%%%%%%%%%%%%%%%%%%%%%%%%%%%%%%%%%%%
\section{Acknowledgements}{Thanks for the support provided by Mindspore Community. Thanks for the valuable suggestions from Ziwen Huang. This work is supported by National Natural Science Foundation of China (Grant Nos. 62371069, 62372048, 62272056 ), and BUPT Excellent Ph.D. Students Foundation(CX2023123).}

%\bibliographystyle{MSP}
%\bibliography{refe}

\newpage
\appendix

% 重置图表编号
\setcounter{figure}{0}
\setcounter{table}{0}
\renewcommand{\thefigure}{\Alph{section}\arabic{figure}} 
\renewcommand{\thetable}{\Alph{section}\arabic{table}}

\begin{appendix}
	%\section{Name}
	% 重置图表编号
	\setcounter{figure}{0}
	\setcounter{table}{0}
	\setcounter{equation}{0}
	\renewcommand{\thefigure}{A\arabic{figure}}
	\renewcommand{\thetable}{A\arabic{table}}
	\renewcommand{\theequation}{A\arabic{equation}}

	\section{\textcolor{black}{Performance analysis of adaptive algorithms with and without the target unitary layer after the first layer}}\label{com}
	%我们在具有 8、10、12 和 14 个顶点的 ER 图（边概率为 0.5）以及三正则图上，探究了自适应类算法在除第一层 Ansatz 外，是否保留目标酉算子 $U(H_C)$ 对算法性能的影响。图中展示了两类算法（Adaptive-QAOA+ 和 AMA-QAOA+）及其变体的表现，其中变体对应保留或丢弃目标酉算子，分别用虚线标注。
	
	%我们从四个指标评估各算法的表现：相同运行次数下的最优近似比（OAR）、平均近似比（AAR）、单次运行中参数优化的平均迭代次数、以及最终构建线路所需的平均 CNOT 门数。
	
	%从 ER 图的结果可以看出，省略目标酉算子后，两种自适应算法的最优近似比都有所提升，同时平均所需迭代次数与 CNOT 门数也减少。其背后的原因在于：省略目标酉算子意味着每一层 Ansatz 中需要优化的参数减少，从而显著缩小了参数空间的维度和复杂度。更小的参数空间一方面降低了优化过程陷入局部最优的风险，使得算法更容易收敛到高质量的解，从而提升近似比；另一方面，也加快了期望函数的收敛速度，减少了每轮优化迭代的步数以及提升了最终构建线路的紧凑性。
	
	%值得注意的是，当 AMA-QAOA+ 保留了交替结构后算法平均性能下降，而 Adaptive-QAOA+ 的平均性能较于其变体平均性能会提升。其原因在于 AMA-QAOA+ 每轮会一次性选择多个混合器构成一层混合酉算子，并共享一个参数 $\beta$。若同时保留目标酉算子层，则每轮需同时优化该层对应的参数 $\gamma$，显著提升了参数之间的耦合度，增加了优化难度。此时若无法有效协调多个参数的共同作用，量子态演化路径可能受阻，导致优化收敛缓慢甚至陷入局部最优。而省略目标酉层后，优化任务简化，有助于加速收敛并提升整体性能。相比之下，Adaptive-QAOA+ 每轮仅添加一个混合器，其局部作用较弱。当省略目标酉算子层后，电路的演化驱动力进一步减弱。若初始参数设置不佳，期望函数变化更易趋于平缓，从而更快触发终止条件，导致平均性能略有下降。综上，省略后续层的目标酉算子对于 AMA-QAOA+ 有助于减轻优化负担、提高解质量与资源效率；而对于 Adaptive-QAOA+，是否保留该结构则需结合具体问题下的性能表现与资源开销进行权衡判断。

\begin{figure*}[htbp]
		\centering
		% 子图1
		\begin{minipage}[b]{0.45\textwidth} % 设置子图宽度为总宽度的45%
			\centering
			\subfloat[ Mean OAR.]{\label{OAR_ER} \includegraphics[width=\textwidth]{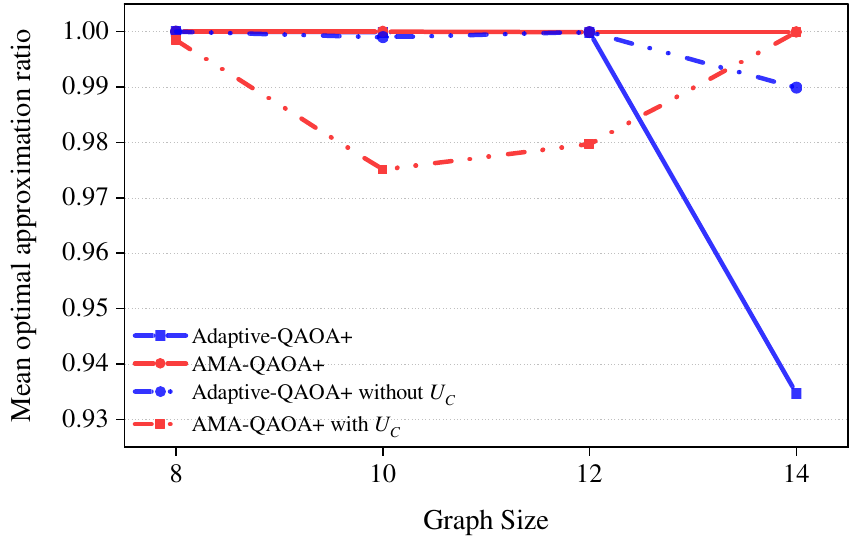}} % 替换为子图的文件名
		\end{minipage}
		\hspace{0.05\textwidth} % 调节间距
		% 子图2
		\begin{minipage}[b]{0.45\textwidth} % 设置子图宽度为总宽度的45%
			\centering
			\subfloat[ Mean AAR.]{\label{OAR_ER} \includegraphics[width=\textwidth]{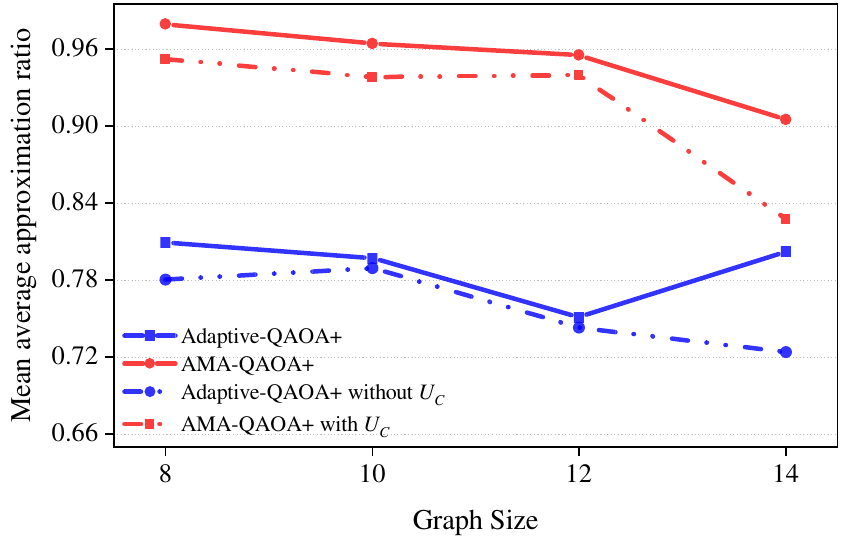}} % 替换为子图的文件名
		\end{minipage}
		
		% 子图3
		\begin{minipage}[b]{0.45\textwidth} % 设置子图宽度为总宽度的45%
			\centering
			\subfloat[ Mean CNOT gates.]{\label{OAR_ER} \includegraphics[width=\textwidth]{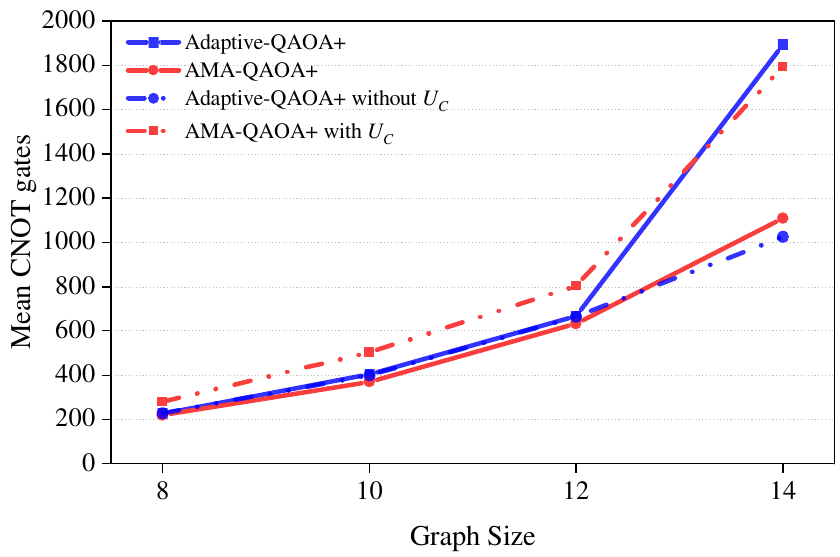}} % 替换为子图的文件名
		\end{minipage}
		\hspace{0.05\textwidth} % 调节间距
		% 子图4
		\begin{minipage}[b]{0.45\textwidth} % 设置子图宽度为总宽度的45%
			\centering
			\subfloat[ Mean Iterations.]{\label{OAR_ER} \includegraphics[width=\textwidth]{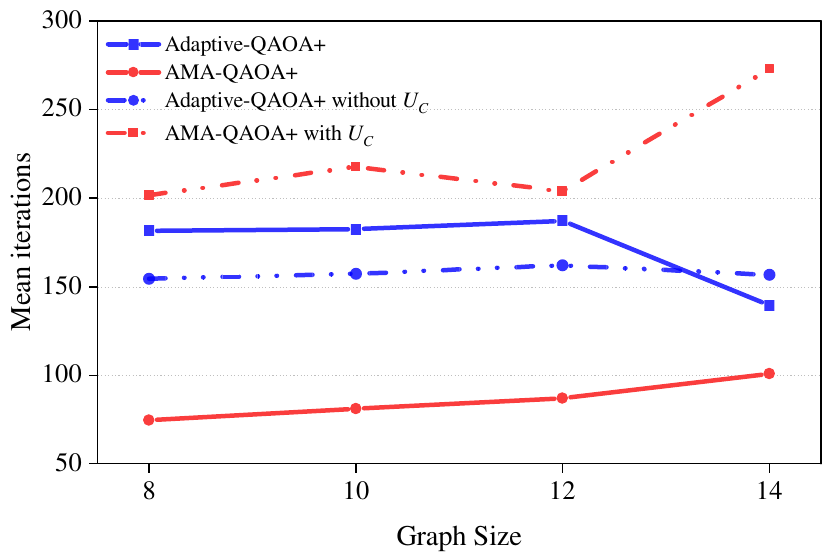}} % 替换为子图的文件名
		\end{minipage}
		\captionsetup{justification=raggedright, singlelinecheck=false}
		\caption{\textcolor{black}{Performance comparison of adaptive algorithms with and without the alternating structure in $i(\ge 2)$-th layer of ansatz on ER graphs of an edge probability of 0.5. Solid lines denote original algorithms, and dashed lines denote variants.}} % 整体图的标题
		\label{com_ER} % 整体图的标签
	\end{figure*}

\begin{figure*}[htbp]
		\centering
		% 子图1
		\begin{minipage}[b]{0.45\textwidth} % 设置子图宽度为总宽度的45%
			\centering
			\subfloat[ Mean OAR.]{\label{OAR_ER} \includegraphics[width=\textwidth]{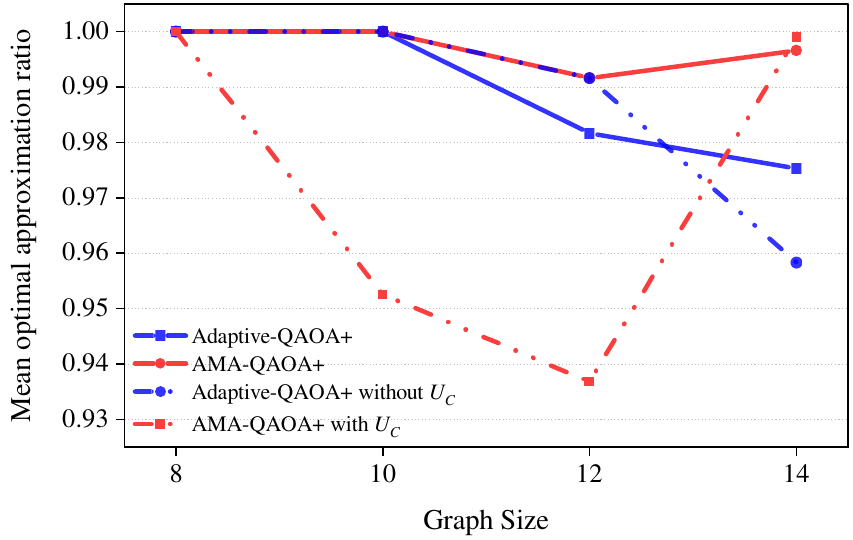}} % 替换为子图的文件名
		\end{minipage}
		\hspace{0.05\textwidth} % 调节间距
		% 子图2
		\begin{minipage}[b]{0.45\textwidth} % 设置子图宽度为总宽度的45%
			\centering
			\subfloat[ Mean AAR.]{\label{OAR_ER} \includegraphics[width=\textwidth]{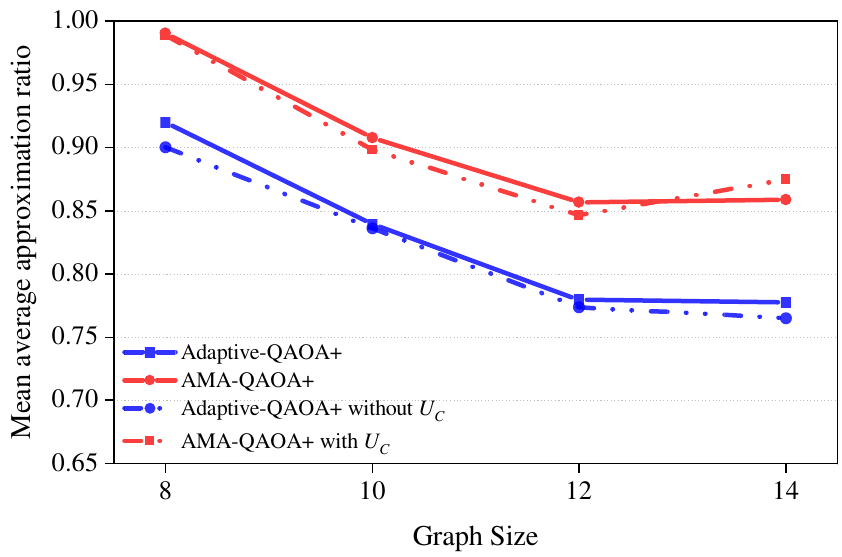}} % 替换为子图的文件名
		\end{minipage}
		
		% 子图3
		\begin{minipage}[b]{0.45\textwidth} % 设置子图宽度为总宽度的45%
			\centering
			\subfloat[ Mean CNOT gates.]{\label{OAR_ER} \includegraphics[width=\textwidth]{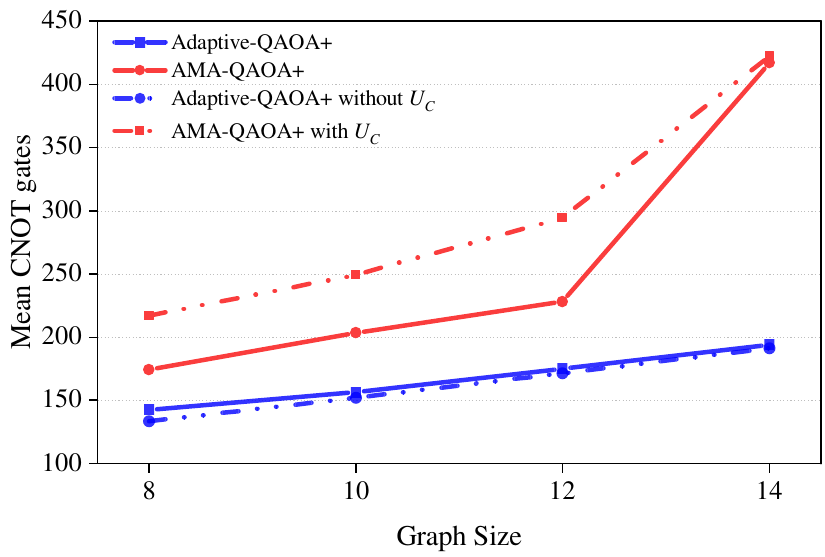}} % 替换为子图的文件名
		\end{minipage}
		\hspace{0.05\textwidth} % 调节间距
		% 子图4
		\begin{minipage}[b]{0.45\textwidth} % 设置子图宽度为总宽度的45%
			\centering
			\subfloat[  Mean Iterations.]{\label{OAR_ER} \includegraphics[width=\textwidth]{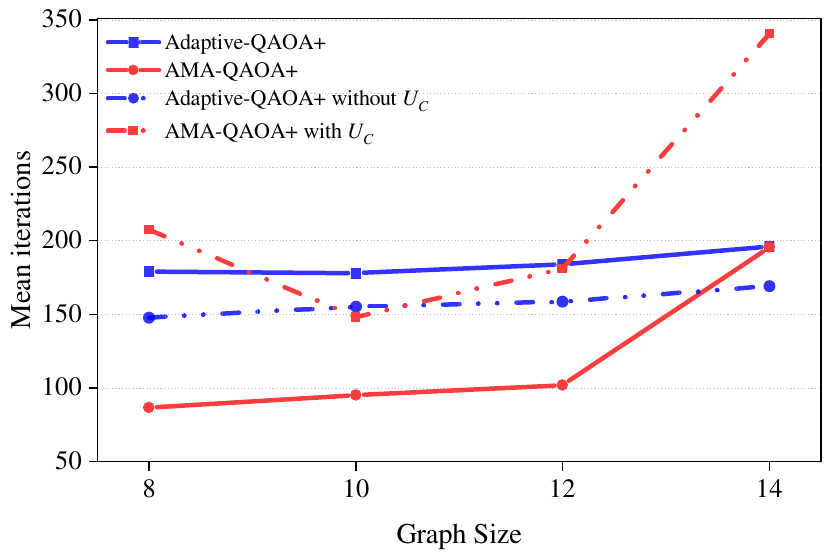}} % 替换为子图的文件名
		\end{minipage}
		\captionsetup{justification=raggedright, singlelinecheck=false}  
		\caption{\textcolor{black}{Performance comparison of adaptive algorithms with and without the alternating structure in $i(\ge 2)$-th layer of ansatz on 3-regular graphs. Solid lines denote original algorithms, and dashed lines denote variants.}} % 整体图的标题
		\label{com_regular} % 整体图的标签
	\end{figure*}

	\textcolor{black}{We investigate the performance impact of retaining or discarding the target unitary operator \( U_C \) after the first layer of the ansatz in adaptive algorithms. The evaluation is conducted on ER graphs with an edge probability of 0.5 and 3-regular graphs, with graph sizes \( n = 8, 10, 12, 14 \). For each graph size, we use 20 ER graph instances and 20 3-regular graph instances. Figure~\ref{com_ER} and Figure~\ref{com_regular} show the performance of Adaptive-QAOA+ and AMA-QAOA+, as well as their corresponding variants that differ in whether the target unitary operator is retained after the first layer of ansatz. These variants are indicated by dashed lines in the figures for visual comparison.}
	
	\medskip
	\textcolor{black}{We evaluate the performance of each algorithm using four metrics under the same number of optimization runs: (1) the mean optimal approximation ratio, (2) the mean average approximation ratio, (3) the mean number of iterations required for parameter optimization per run, and (4) the mean number of CNOT gates required to implement the final parameterized quantum circuit.
	}
	
	\medskip
	\textcolor{black}{From the results in Figure~\ref{com_ER}, we observe that discarding the target unitary operator improves the mean optimal approximation ratio for both adaptive algorithms on ER graphs. Meanwhile, it also reduces the mean number of optimization iterations and the number of CNOT gates required per run. This performance improvement can be primarily attributed to the reduced number of optimized parameters in each layer, which significantly lowers the dimensionality and complexity of the parameter space. A smaller parameter space reduces the risk of getting trapped in local optima, enabling the algorithm to converge more easily to high-quality solutions and thus improving the approximation ratio. In addition, the reduction in parameters shortens the number of iterations required in each optimization round, leading to an overall decrease in iterations.
	}
	
	\medskip
	\textcolor{black}{It is worth noting that the two adaptive algorithms exhibit different behaviors when retaining the alternating structure. Specifically, the variant of AMA-QAOA+ that retains the alternating structure shows a decrease in average performance \textcolor{black}{compared with} the version that discards it, whereas Adaptive-QAOA+ shows a slight improvement in average performance relative to its variant without the alternating structure. This behavior arises because AMA-QAOA+ selects multiple mixers in each round, which together form a mixer unitary layer that shares a common parameter $\beta_{i}$. Finding a single \(\beta_i\) that simultaneously drives multiple mixers effectively is already a challenging task. When the target unitary operator is retained, the additional parameter \(\gamma_i\) further increases the parameter coupling and exacerbates the optimization difficulty, leading to slower convergence or even getting trapped in low-quality solutions. By contrast, discarding the target unitary operator simplifies the optimization task, accelerates convergence, and enhances overall performance. In comparison, Adaptive-QAOA+ adds only one mixer operator in each round, and its local impact may be relatively small. When the target unitary operator is omitted, the overall driving effect of the circuit on quantum state evolution is further weakened. As a result, if the initial parameters are poorly chosen, the expectation function may vary more slowly, increasing the likelihood of triggering stop conditions, which slightly reduces the average performance but may also lead to fewer optimization iterations and lower CNOT gate costs per run. The above analysis also applies to the consistent performance trends observed on 3-regular graphs, which are omitted here for brevity.
	}
	
	\medskip
	\textcolor{black}{In summary, omitting the target unitary operator in later layers proves beneficial for AMA-QAOA+, as it alleviates optimization complexity and enhances both solution quality and resource efficiency. In the case of Adaptive-QAOA+, however, the impact is less clear-cut, and the decision to retain or discard the alternating structure should be guided by a trade-off between performance and resource usage under different problem settings.}

	\section{\textcolor{black}{Calculating average initial expectation function values and gradients}} \label{calculate_F_grad}
	
	\textcolor{black}{To guide the adaptive mixer selection in AMA-QAOA+, the average initial expectation function value and the average gradient are evaluated for each candidate mixer. Suppose that the variational circuit has already been optimized up to layer $i{-}1$ ($i \geq 2$), with the optimized parameters denoted by $\gamma_{1}^{*}$ and $\boldsymbol{\beta}_{1:i-1}^{*}$. Then, the corresponding output quantum state at this stage is given by}
	\begin{equation}
		\textcolor{black}{|\psi_{i-1}(\gamma_1^{*}, \boldsymbol{\beta}_{1:i-1}^{*})\rangle = \prod_{l=1}^{i-1} \mathrm{e}^{-\mathrm{i} \beta_l^{*} H_{M}^{l}} \mathrm{e}^{-\mathrm{i} \gamma_1^{*} H_C} |s\rangle,}
		\label{eq:current_output_state}
	\end{equation}
	\textcolor{black}{where \( H_{M}^{l} \) denotes the mixer Hamiltonian used in the \( l \)-th layer of the mixer unitary operator, which is composed of several partial mixer Hamiltonians.}
	
	\medskip
	\textcolor{black}{To assess which candidate mixer operator can be added to the $i$-th mixer layer, for each candidate mixer, a trial circuit is constructed by appending the corresponding unitary \( U_{M_j}(\beta_{i}) \) to the existing optimized circuit. The initial parameters of the optimized layers are set to their previously optimized values \( \boldsymbol{\gamma}_{1}^{*} \) and \( \boldsymbol{\beta}_{1:i-1}^{*} \), while the newly introduced parameter \( \beta_i \) associated with the $i$-th layer mixer is randomly initialized. Under this initial parameter configuration, the output quantum state becomes}
	\begin{equation}
		\textcolor{black}{|\psi_i(\gamma_{1}^{*}, \boldsymbol{\beta}_{1:i-1}^{*}, \beta_i)\rangle = U_{M_j}(\beta_i) |\psi_{i-1}(\gamma_{1}^{*},\boldsymbol{\beta}_{1:i-1}^{*})\rangle.}
	\end{equation}
	
	\textcolor{black}{The corresponding initial expectation function value of this state with respect to the target Hamiltonian \( H_C \) is given by}
	\begin{equation}
		\textcolor{black}{F_{\text{init}}(\gamma_{1}^{*}, \boldsymbol{\beta}_{1:i-1}^{*}, \beta_i) = \langle \psi_i(\gamma_{1}^{*}, \boldsymbol{\beta}_{1:i-1}^{*}, \beta_i) | H_C | \psi_i(\gamma_{1}^{*}, \boldsymbol{\beta}_{1:i-1}^{*}, \beta_i)\rangle.}
	\end{equation}
	
	\textcolor{black}{By averaging the initial expectation function values under $N$ different sets of random initial parameters, the average initial expectation function value can be obtained for this candidate mixer.}
	%\begin{equation}
	%\textcolor{black}{F_{\text{fun}} = \frac{1}{N} \sum_{j=1}^{N} F_{\text{init}}^{(j)}}
	%\end{equation}
	%\textcolor{black}{This value reflects the quality of the solution that may be achieved by adding the candidate mixer without further parameter optimization. A lower initial expectation value often indicates that the circuit with the added mixer tends to produce better-quality states even before optimization. Such mixers are more likely to accelerate convergence or help the algorithm escape poor local minima. }
	
	\medskip
	To evaluate the trainability of each candidate mixer, we calculate the gradient of the expectation function $\langle H_C \rangle$ with respect to the newly introduced parameter $\beta_{i}$ using the parameter-shift rule. \textcolor{black}{When the parameter $\beta_{i}$ is shared by multiple partial mixer unitaries $U_{M_j}(\beta_i)$ within a single layer, the gradient of $\langle H_C \rangle$ with respect to the shared parameter $\beta_{i}$ is decomposed into the sum of derivatives of the expected function $\langle H_C \rangle$ with respect to each parameter $\beta_{i,j}$, where the parameter $\beta_{i,j}$ is related to the $j$-th partial mixer in the $i$-th layer of the mixer unitary operator. Specifically, the above gradient can be calculated by}
	\begin{equation}
		\textcolor{black}{\frac{\partial \langle H_C \rangle}{\partial \beta_{i}} = \sum_{j \in S} \frac{\partial \langle H_C \rangle}{\partial \beta_{i,j}},}
	\end{equation}
	\textcolor{black}{where $S$ is the set of all partial mixers sharing $\beta_{i}$.}
	
	\medskip
	\textcolor{black}{The gradient of the expected function with respect to the parameter $\beta_{i,j}$ is individually calculated by}
	\begin{equation}
		\textcolor{black}{\frac{\partial \langle H_C \rangle}{\partial \beta_{i,j}}= \frac{1}{2} \left[ \langle H_{C} \rangle_{\beta_{i,j} + \frac{\pi}{2}} - \langle H_{C} \rangle_{\beta_{i,j} - \frac{\pi}{2}} \right].}
	\end{equation}
	\textcolor{black}{The \( \langle H_C \rangle_{\beta_{i,j} \pm \frac{\pi}{2}} \) denotes the expectation value of the target Hamiltonian with respect to the output quantum state that is prepared with the following parameter configuration. Specifically, the shift of $\pm \frac{\pi}{2}$ is applied \textbf{only} to the parameter $\beta_{i,j}$, and all other parameters in the circuit (including other parameters that are related to the other partial mixer in the set $S$, $\gamma_{1}^{*}$ and the full set of $\boldsymbol{\beta}_{1:i}^{*}$) are held at their original, unshifted value. }
	
	\iffalse
	\textcolor{black}{To evaluate the trainability of each candidate mixer, we calculate the gradient of the expectation function $\langle H_{C} \rangle$ with respect to the newly introduced parameter \( \beta_i \) using the parameter-shift rule. Specifically, the gradient under one set of random initial parameters is given by}
	\begin{equation}
		\textcolor{black}{\frac{\partial F }{\partial \beta_i} = \frac{1}{2} \left[ \langle H_{C} \rangle_{\beta_i + \frac{\pi}{2}} - \langle H_{C} \rangle_{\beta_i - \frac{\pi}{2}} \right],}
	\end{equation}
	\textcolor{black}{where \( \langle H_C \rangle_{\beta_i \pm \frac{\pi}{2}} \) denotes the expectation value of the target Hamiltonian evaluated on the quantum state with the \( i \)-th layer parameter shifted by \( \pm \frac{\pi}{2} \), while other parameters are set $(\gamma_{1}^{*}, \boldsymbol{\beta}_{1:i-1}^{*})$.}
	\fi

	\section{\textcolor{black}{Analysis of the impact of incorporating the $F_{\text{fun}}$ into the evaluation function on the performance of AMA-QAOA+}} \label{F_fun_effect}

	\textcolor{black}{To investigate whether incorporating the average initial expectation function value into the evaluation function can improve the solution quality of AMA-QAOA+ or reduce the number of iterations and the CNOT gate count required per run, we designed and conducted the following experiments.}
	
\begin{figure*}[htbp]
		\centering
		% 子图1
		\begin{minipage}[b]{0.95\textwidth} % 设置子图宽度为总宽度的45%
			\centering
			\subfloat[$n=8$, ER graphs.]{\label{n8_ER} \includegraphics[width=\textwidth]{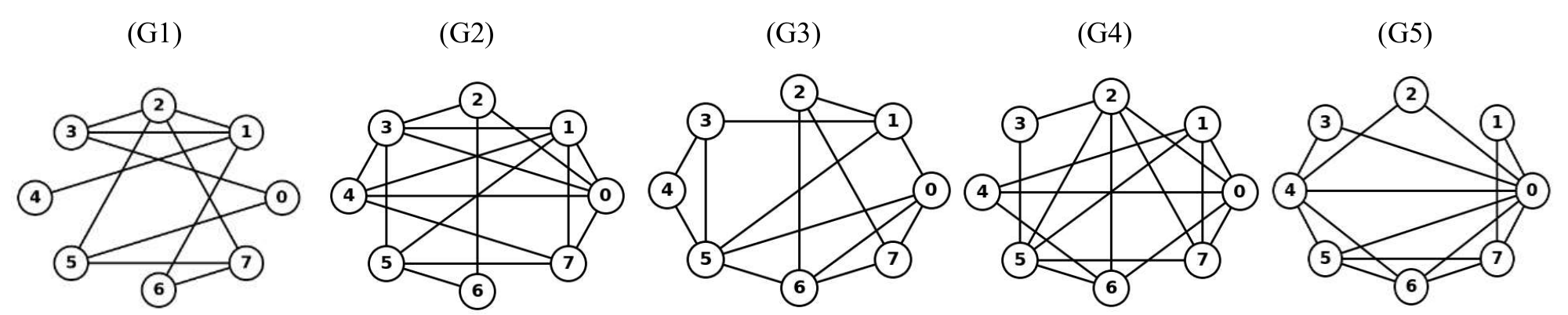}} % 替换为子图的文件名
		\end{minipage}
		\hspace{0.05\textwidth} % 调节间距

		\begin{minipage}[b]{0.95\textwidth} % 设置子图宽度为总宽度的45%
			\centering
			\subfloat[$n = 8$, 3-regular graphs.]{\label{n8_regular} \includegraphics[width=\textwidth]{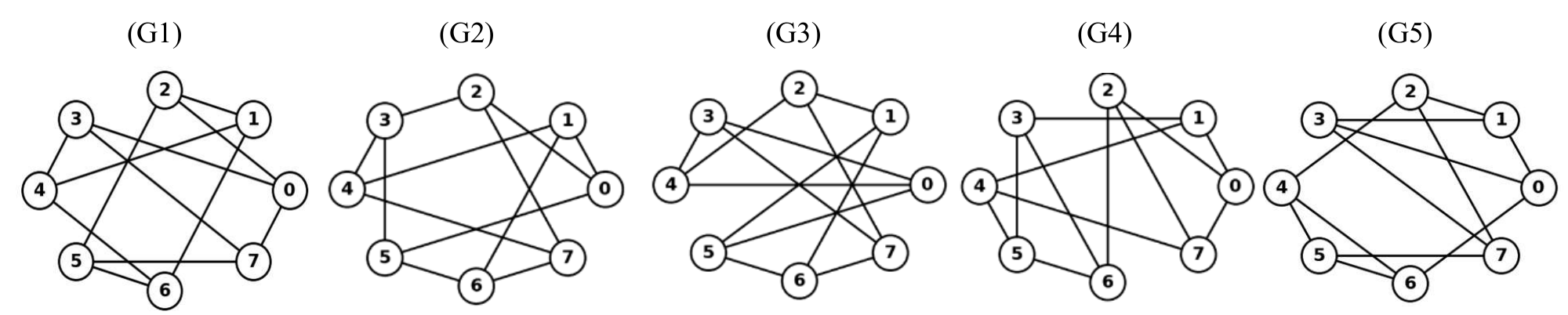}} % 替换为子图的文件名
		\end{minipage}
		\captionsetup{justification=raggedright, singlelinecheck=false}  
		\caption{\textcolor{black}{Five ER graphs with $n = 8$ and edge probability 0.5, and five 3-regular graphs with $n = 8$. These graphs are used to evaluate the performance of AMA-QAOA+ under different evaluation settings.}} % 整体图的标题
		\label{graph_n8} % 整体图的标签
	\end{figure*}

\begin{figure*}[htbp]
		\centering
		% 子图1
		\begin{minipage}[b]{0.435\textwidth} % 设置子图宽度为总宽度的45%
			\centering
			\subfloat[ OAR and AAR, ER graphs.]{\label{AR_ER_n8_F_fun} \includegraphics[width=\textwidth]{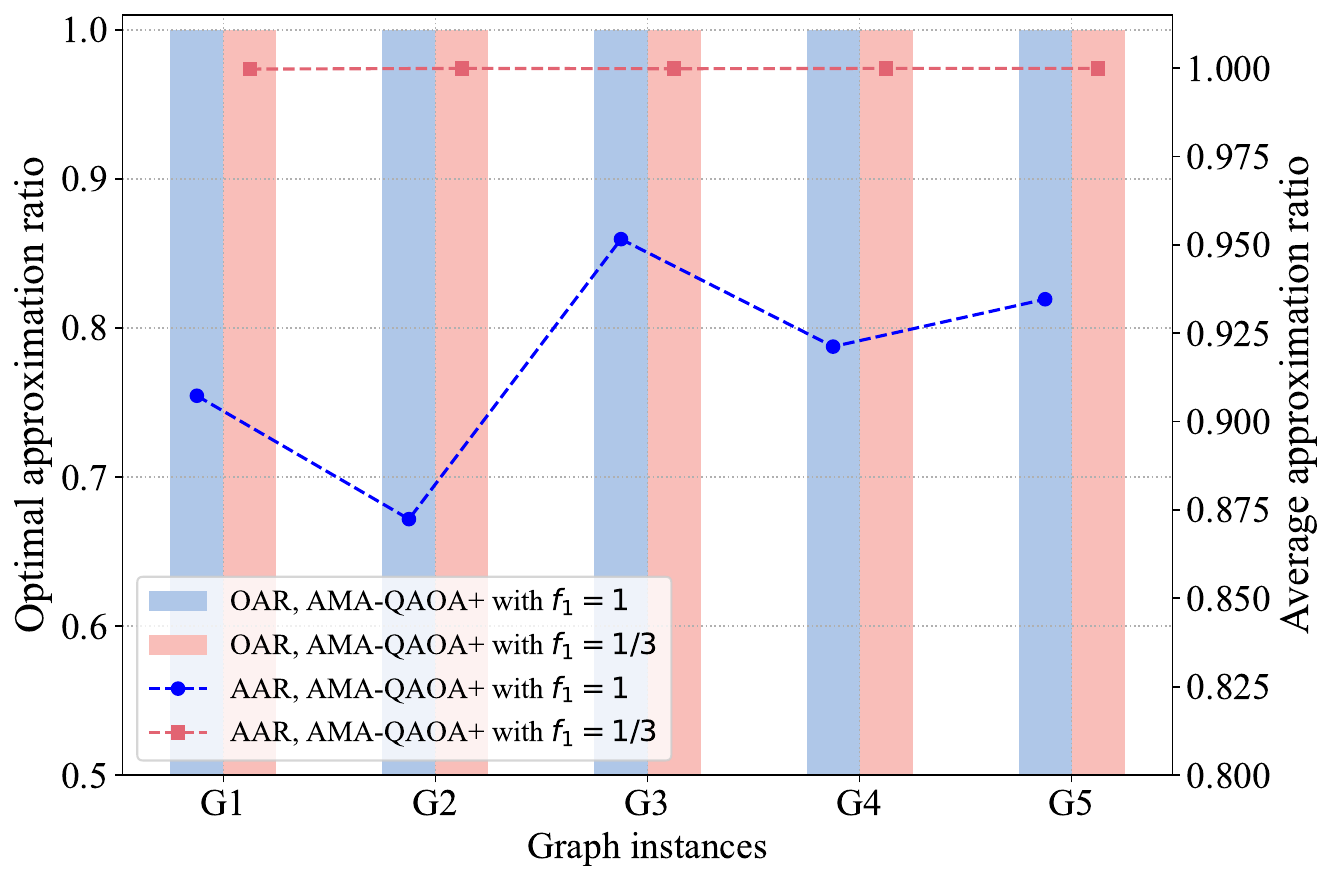}} % 替换为子图的文件名
		\end{minipage}
		\hspace{0.05\textwidth} % 调节间距
		% 子图2
		\begin{minipage}[b]{0.435\textwidth} % 设置子图宽度为总宽度的45%
			\centering
			\subfloat[ Average iterations and CNOT gates, ER graphs.]{\label{R_ER_n8_F_fun} \includegraphics[width=\textwidth]{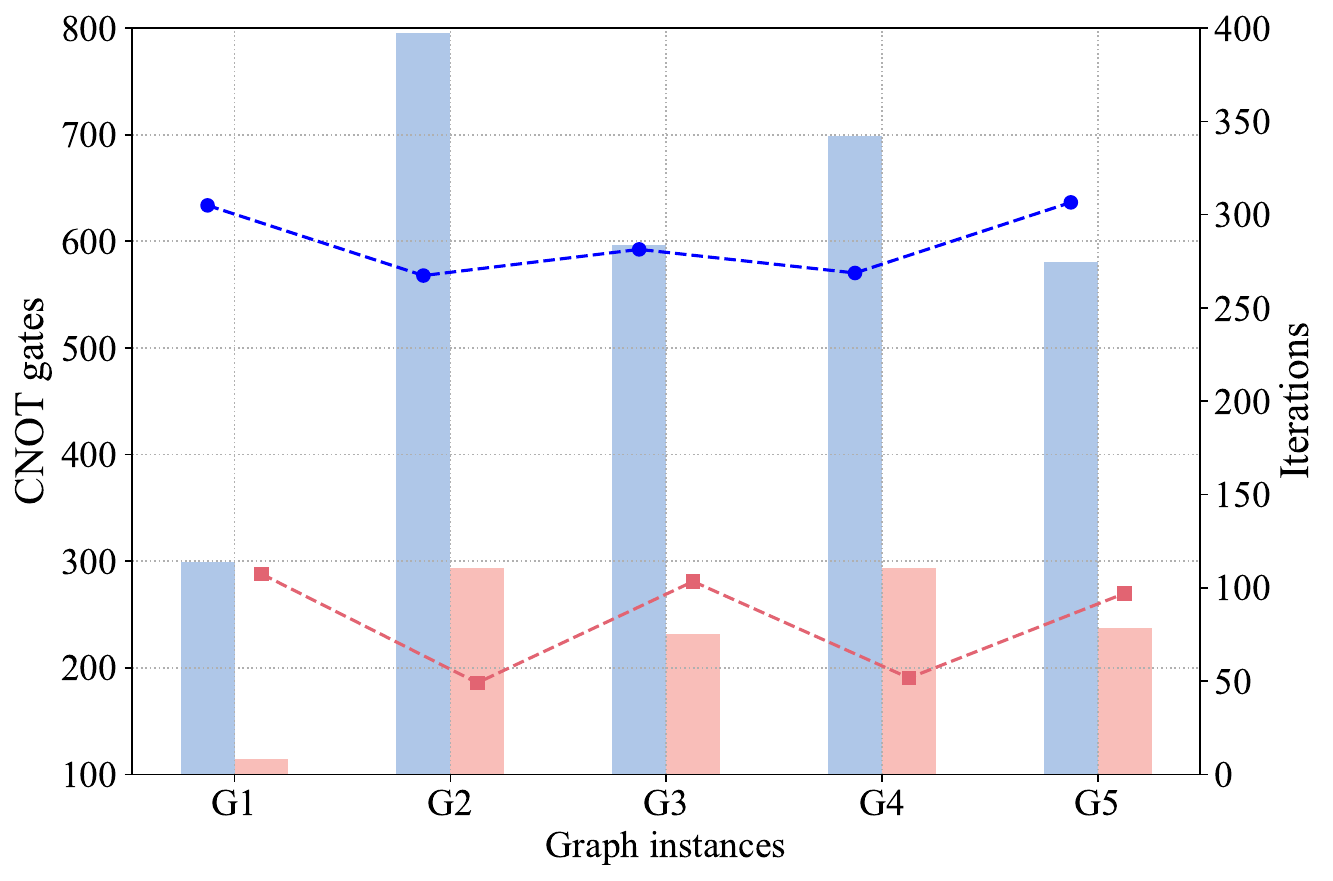}} % 替换为子图的文件名
		\end{minipage}
		
		% 子图3
		\begin{minipage}[b]{0.435\textwidth} % 设置子图宽度为总宽度的45%
			\centering
			\subfloat[ OAR and AAR, 3-regular graphs.]{\label{AR_regular_n8_F_fun} \includegraphics[width=\textwidth]{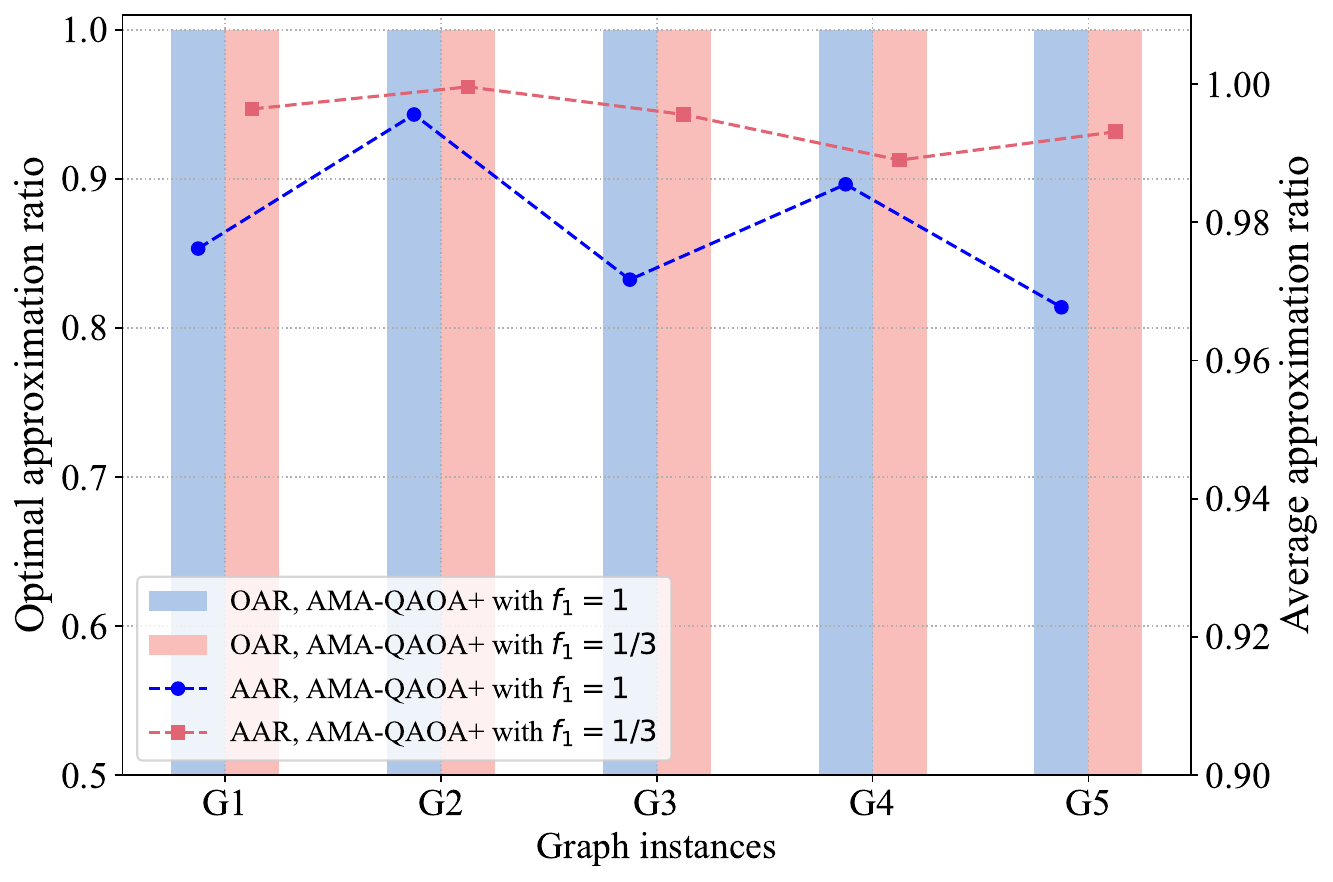}} % 替换为子图的文件名
		\end{minipage}
		\hspace{0.05\textwidth} % 调节间距
		% 子图4
		\begin{minipage}[b]{0.435\textwidth} % 设置子图宽度为总宽度的45%
			\centering
			\subfloat[ Average iterations and CNOT gates, 3-regular graphs.]{\label{R_regular_n8_F_fun} \includegraphics[width=\textwidth]{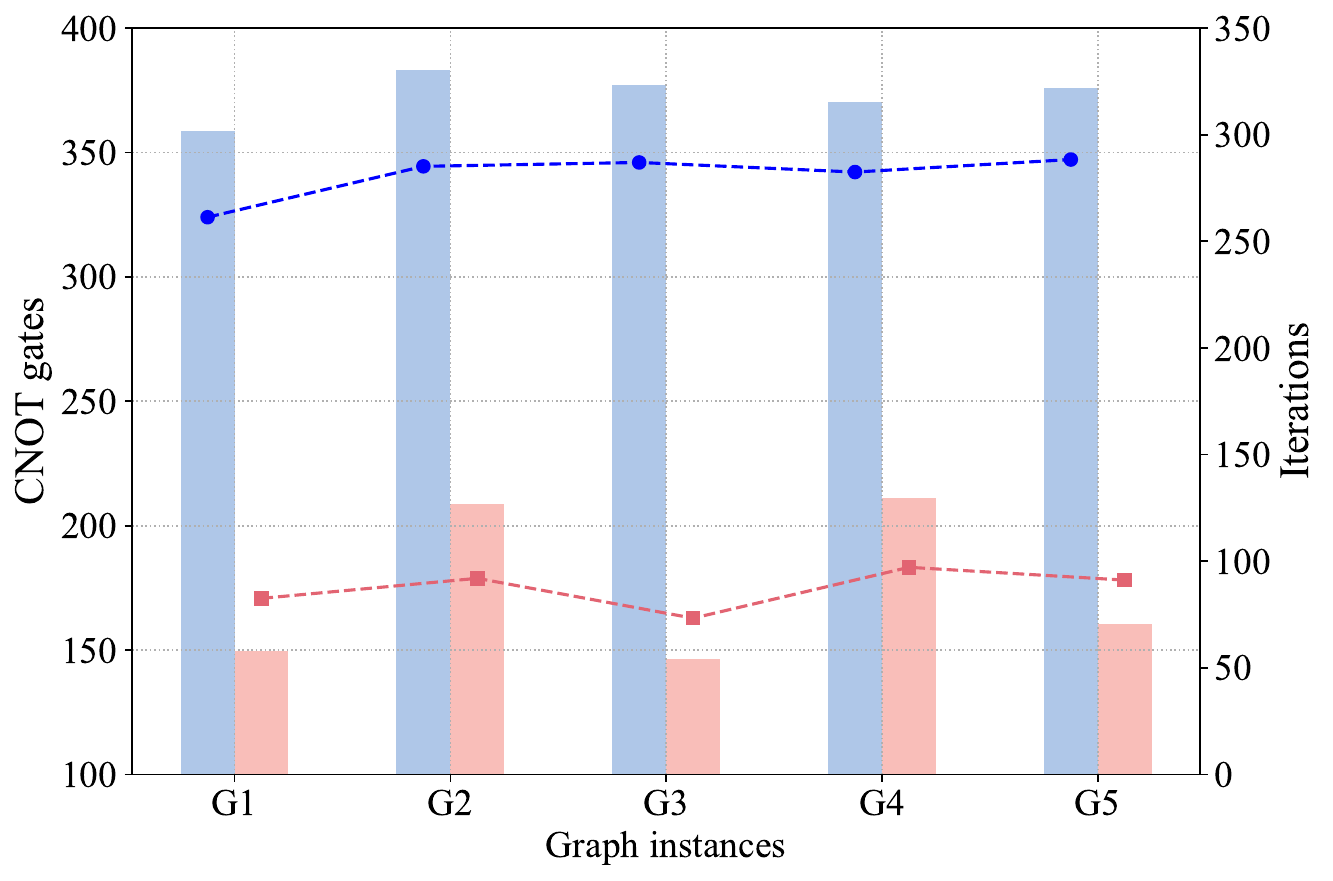}} % 替换为子图的文件名
		\end{minipage}
		\captionsetup{justification=raggedright, singlelinecheck=false}  
		\caption{\textcolor{black}{Performance comparison of AMA-QAOA+ on graphs with $n = 8$ under two evaluation settings, where $f_1 = 1$ corresponds to using only the average gradient in the evaluation function, and $f_1 = \frac{1}{3}$ corresponds to using both the average gradient and the average initial expectation function value. 
		}} % 整体图的标题
		\label{F_fun_effect_n8} % 整体图的标签
	\end{figure*}

\begin{figure*}[htbp]
		\centering
		% 子图1
		\begin{minipage}[b]{0.95\textwidth} % 设置子图宽度为总宽度的45%
			\centering
			\subfloat[$n=14$, ER graphs.]{\label{n14_ER} \includegraphics[width=\textwidth]{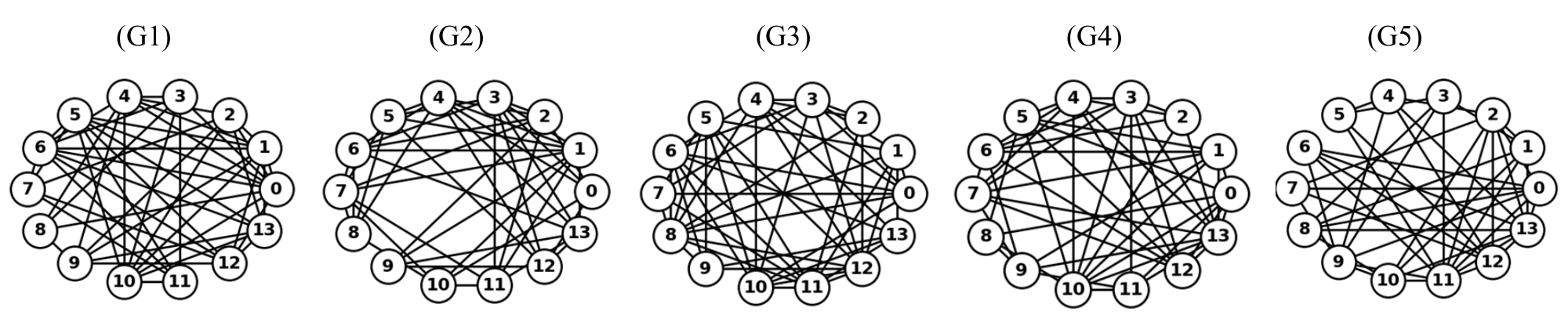}} % 替换为子图的文件名
		\end{minipage}
		\hspace{0.05\textwidth} % 调节间距

		\begin{minipage}[b]{0.95\textwidth} % 设置子图宽度为总宽度的45%
			\centering
			\subfloat[$n = 14$, 3-regular graphs.]{\label{n14_regular} \includegraphics[width=\textwidth]{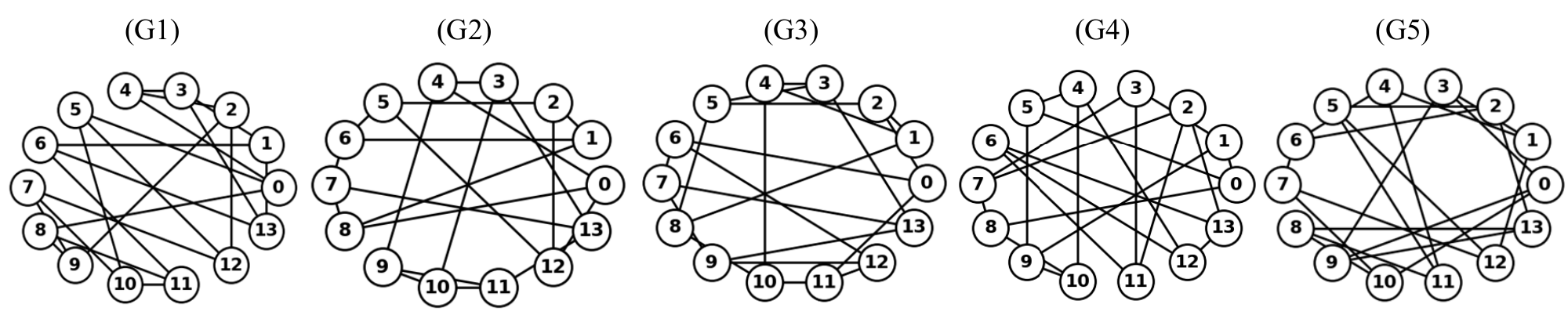}} % 替换为子图的文件名
		\end{minipage}
		\captionsetup{justification=raggedright, singlelinecheck=false}  
		\caption{\textcolor{black}{Five ER graphs with $n = 14$ and edge probability 0.5, and five 3-regular graphs with $n = 14$. These graphs are used to evaluate the performance of AMA-QAOA+ under different evaluation settings.
		}} % 整体图的标题
		\label{graph_n14} % 整体图的标签
	\end{figure*}

\begin{figure*}[htbp]
		\centering
		% 子图1
		\begin{minipage}[b]{0.435\textwidth} % 设置子图宽度为总宽度的45%
			\centering
			\subfloat[ OAR and AAR, ER graphs.]{\label{AR_ER_n14_F_fun} \includegraphics[width=\textwidth]{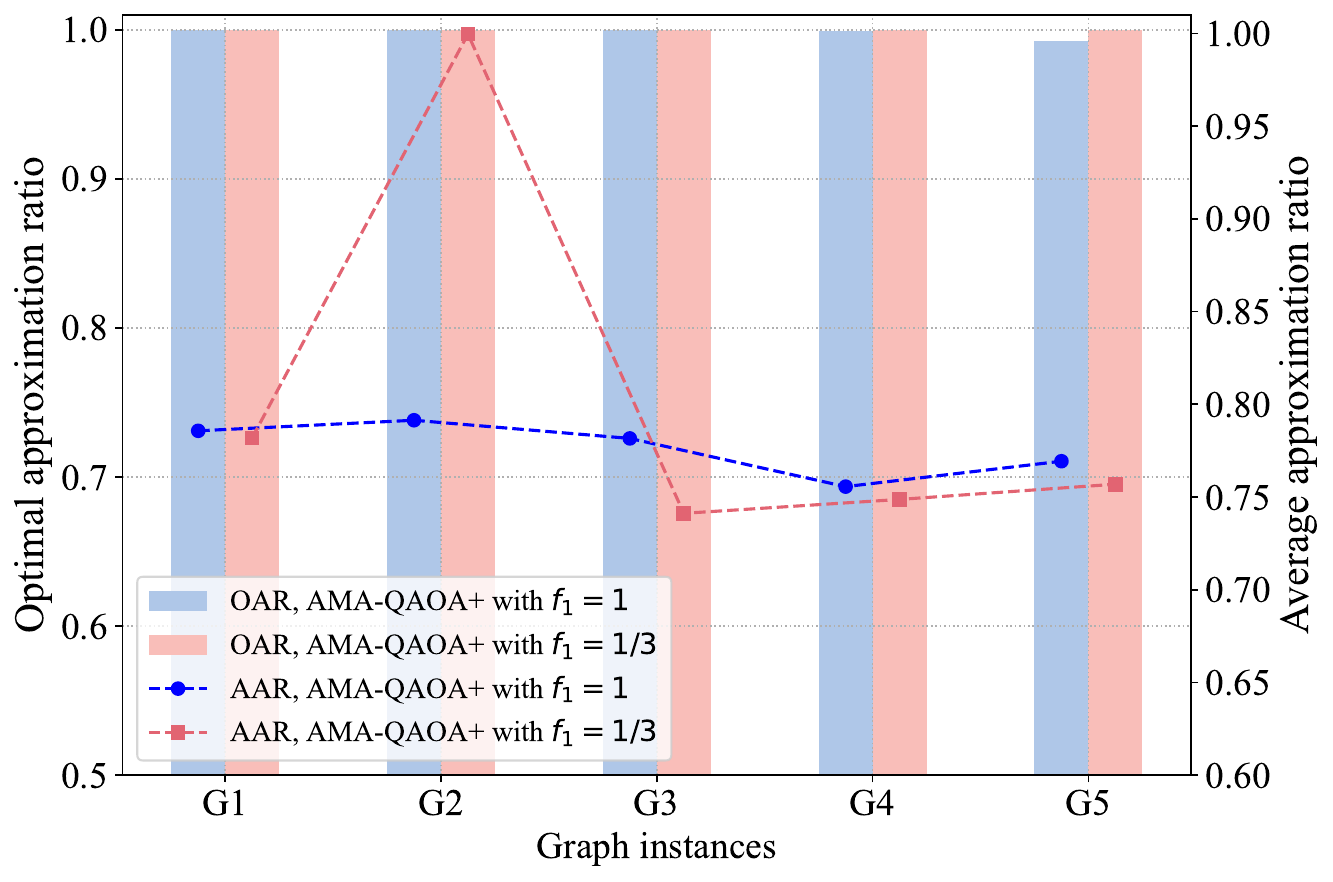}} % 替换为子图的文件名
		\end{minipage}
		\hspace{0.05\textwidth} % 调节间距
		% 子图2
		\begin{minipage}[b]{0.435\textwidth} % 设置子图宽度为总宽度的45%
			\centering
			\subfloat[ Average iterations and CNOT gates, ER graphs.]{\label{R_ER_n14_F_fun} \includegraphics[width=\textwidth]{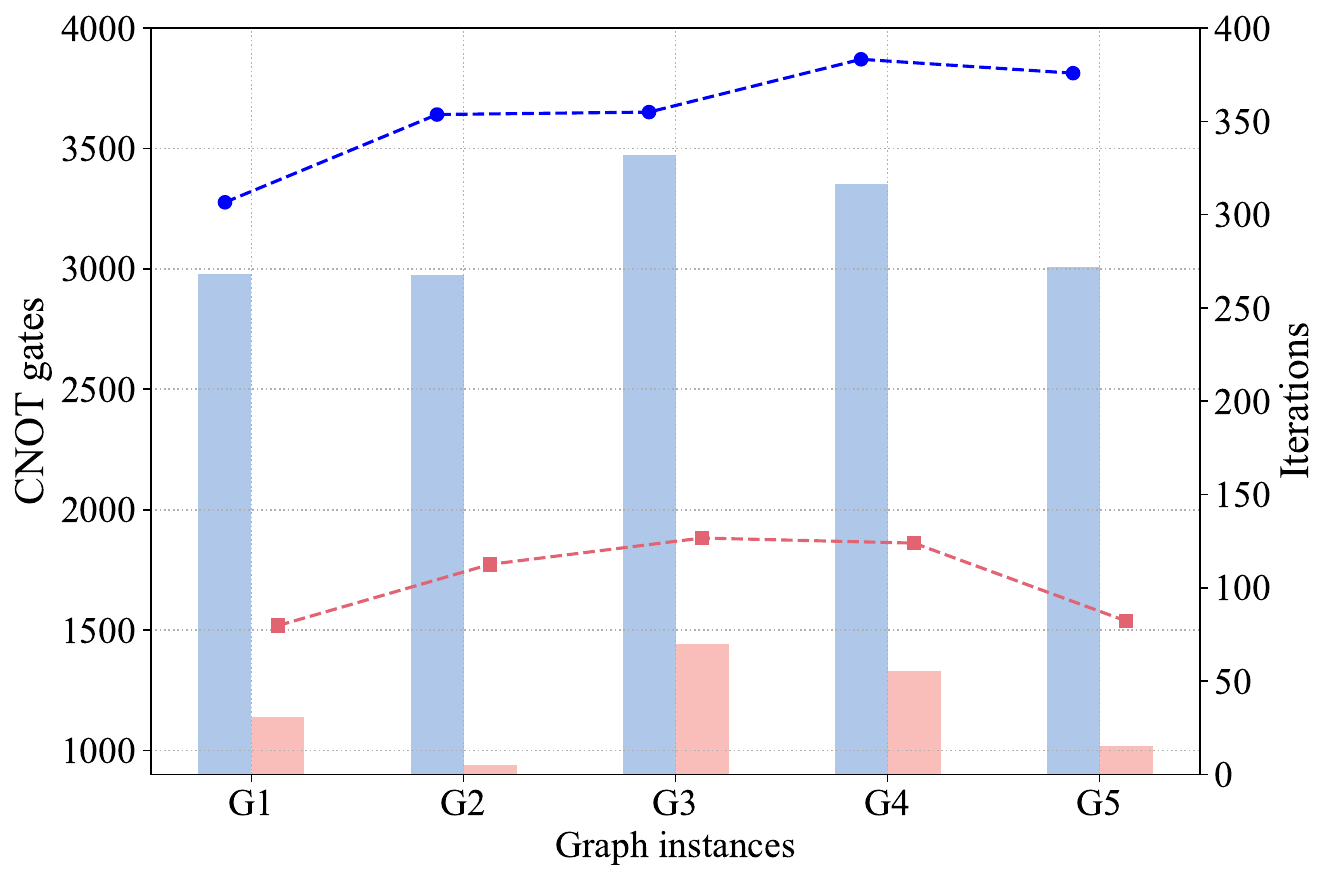}} % 替换为子图的文件名
		\end{minipage}
		
		% 子图3
		\begin{minipage}[b]{0.435\textwidth} % 设置子图宽度为总宽度的45%
			\centering
			\subfloat[ OAR and AAR, 3-regular graphs.]{\label{AR_regular_n14_F_fun} \includegraphics[width=\textwidth]{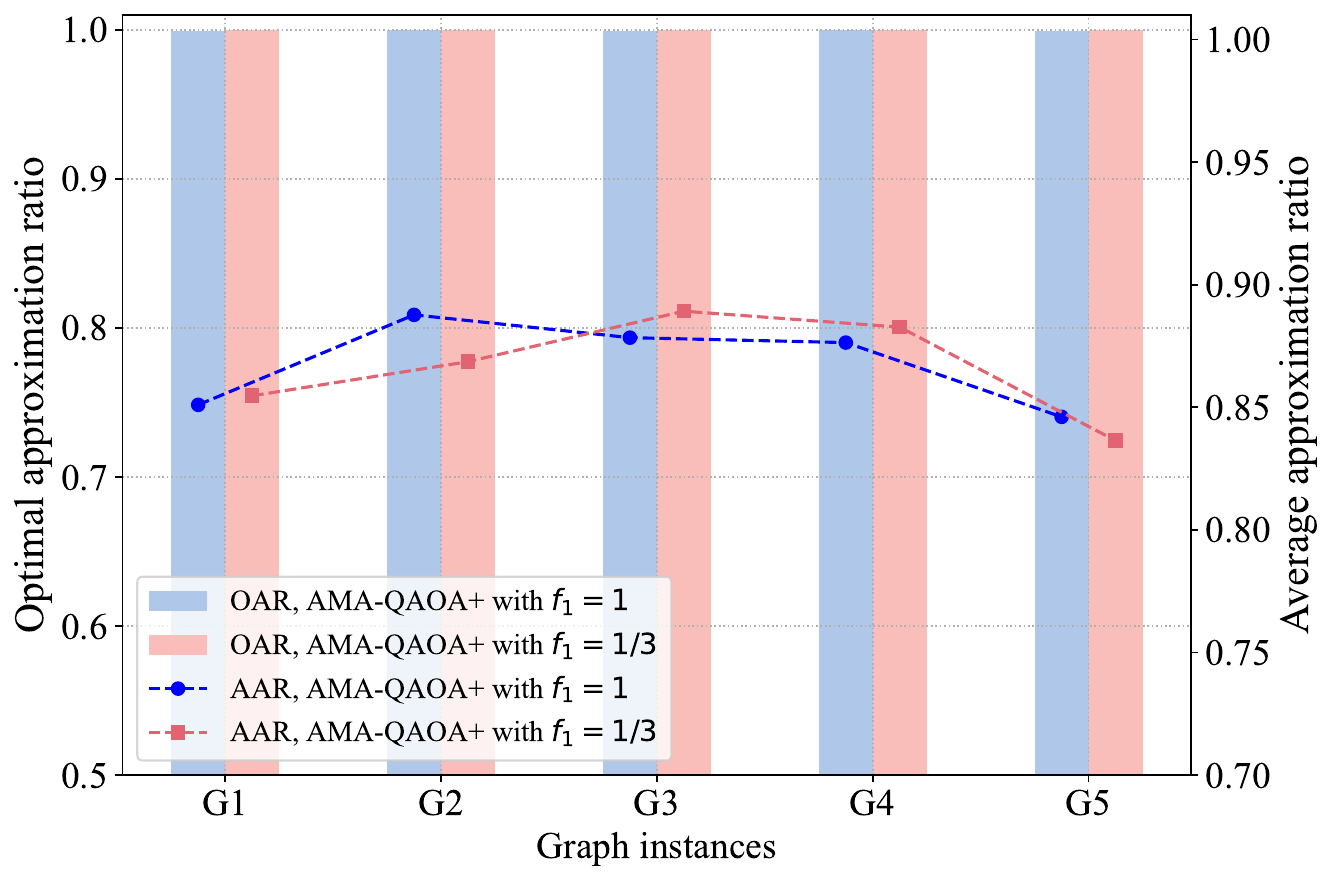}} % 替换为子图的文件名
		\end{minipage}
		\hspace{0.05\textwidth} % 调节间距
		% 子图4
		\begin{minipage}[b]{0.435\textwidth} % 设置子图宽度为总宽度的45%
			\centering
			\subfloat[ Average iterations and CNOT gates, 3-regular graphs.]{\label{R_regular_n14_F_fun} \includegraphics[width=\textwidth]{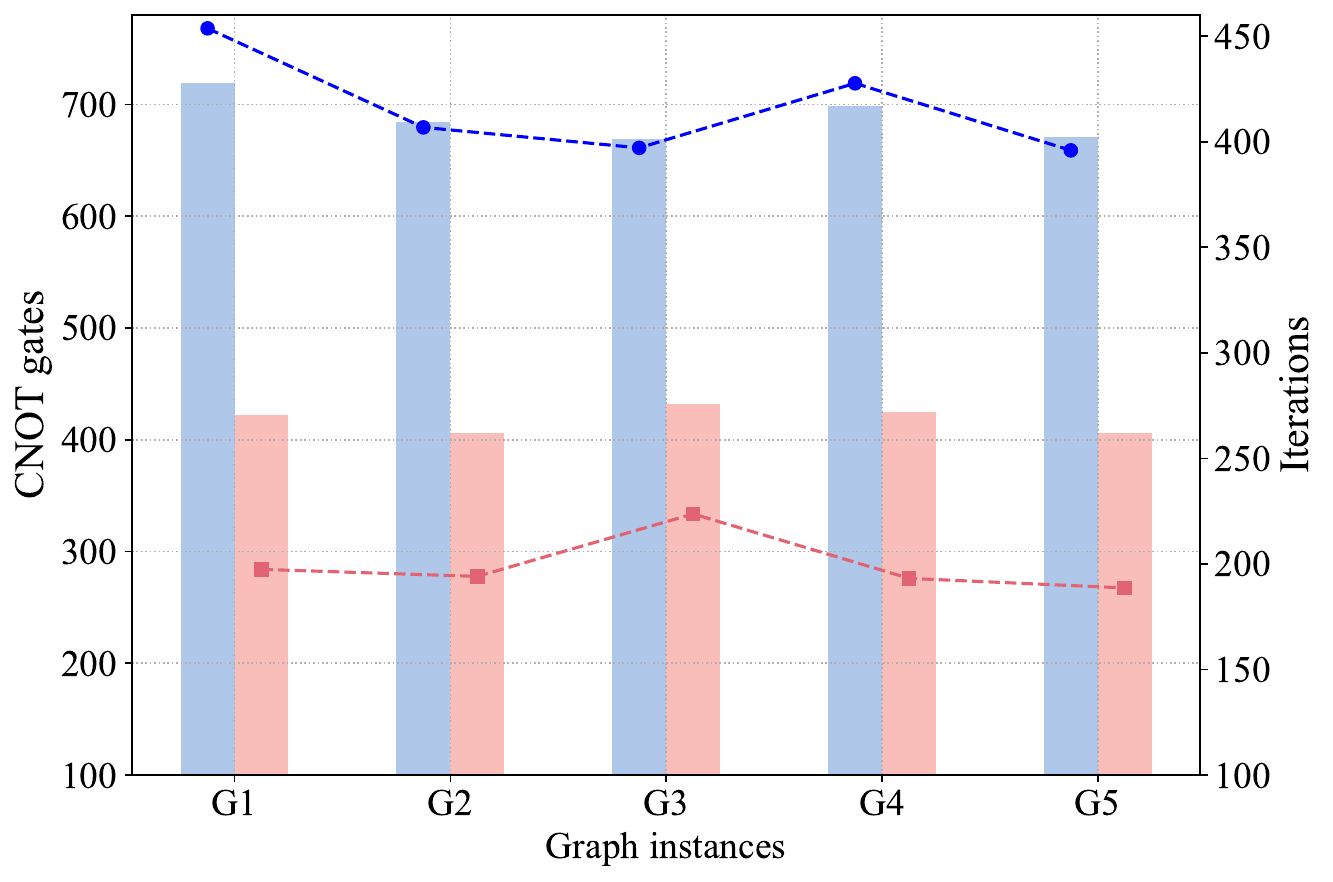}} % 替换为子图的文件名
		\end{minipage}
		\captionsetup{justification=raggedright, singlelinecheck=false}  
		\caption{\textcolor{black}{Performance comparison of AMA-QAOA+ on graphs with $n = 14$ under two evaluation settings, where $f_1 = 1$ corresponds to using only the average gradient in the evaluation function, and $f_1 = \frac{1}{3}$ corresponds to using both the average gradient and the average initial expectation function value.
		}} % 整体图的标题
		\label{F_fun_effect_n14} % 整体图的标签
	\end{figure*}

	\medskip
	\textcolor{black}{Specifically, we randomly selected five ER graphs and five 3-regular graphs with 8 vertices, five ER graphs and five 3-regular graphs with 14 vertices from a total of 160 tested instances. The selected graphs are shown in Figure~\ref{graph_n8} and Figure~\ref{graph_n14}. For these instances, we compared two configurations of the AMA-QAOA+ algorithm. One configuration uses an evaluation function that considers only the average gradient information with \(f_1 = 1\), while the other considers both the average gradient and the average initial expectation function value with \(f_1 = 1/3\). On each instance, both configurations of AMA-QAOA+ were executed independently for 100 optimization runs. In each evaluation, the average gradient and the average initial expectation value were computed using 10 sets of randomly initialized parameters. We recorded four evaluation metrics for each configuration, namely the optimal approximation ratio, the average approximation ratio, the average number of iterations required for parameter optimization per run, and the average number of CNOT gates in the final constructed circuit. The numerical results are presented in Figure~\ref{F_fun_effect_n8} and Figure~\ref{F_fun_effect_n14}, respectively.
	}
	
	\medskip
	\textcolor{black}{Figure~\ref{AR_ER_n8_F_fun} and Figure~\ref{R_ER_n8_F_fun} show the performance comparison of the two versions of AMA-QAOA+ on the 8-vertex ER graphs, while Figure~\ref{AR_regular_n8_F_fun} and Figure~\ref{R_regular_n8_F_fun} present the results for the 8-vertex 3-regular graphs. In each subfigure, the bar charts display the OAR and the average number of CNOT gates, while the line plots present the AAR and the average number of iterations. Blue corresponds to the configuration with \(f_1 = 1\), which excludes the average initial expectation value in the evaluation function. Red corresponds to the configuration with \(f_1 = 1/3\), which incorporates the average initial expectation function value. From the results, it can be observed that incorporating the average initial expectation function value can improve the average approximation ratio across all given instances. Moreover, it also significantly reduces the average number of iterations and CNOT gates. The underlying reason is that the average initial expectation function value reflects the quality of the initial solution after adding a candidate mixer before any parameter optimization is performed. A configuration with a higher-quality initial solution is more likely to approach a high-quality quasi-optimal solution efficiently during the optimization process, thereby reducing the required iterations and producing a more compact circuit. In contrast, when the initial solution is of lower quality, the number of iterations required to reach convergence within a single optimization round may increase. It may also lead to convergence to a poor local optimum. In such cases, the expectation function value may oscillate among different optimization rounds, making it harder for the algorithm to satisfy the stopping condition \textcolor{black}{given in Equation~\eqref{eq:stopping_condition}}. This may increase the number of optimization rounds in a single run and lead to higher overall resource consumption in terms of both iterations and CNOT gates.
	}
	
	\medskip
	\textcolor{black}{We also conducted similar experiments on selected 14-vertex ER and 3-regular graph instances. The results, as shown in Figure~\ref{F_fun_effect_n14}, show that incorporating the average initial expectation function value into the evaluation function does not consistently improve the approximation ratios. However, it still leads to a significant reduction in the average number of optimization iterations and CNOT gates per run. This phenomenon can be attributed to the increased difficulty posed by larger problem sizes. As the number of vertices increases, the solution space expands exponentially, making it much more difficult to identify initial parameter configurations that are close to the quasi-optimal solution. In this high-dimensional space, a mixer with a relatively low average initial expectation value may only offer locally favorable behavior without guaranteeing the circuit toward global optimal solutions. Nevertheless, \textcolor{black}{compared with} evaluation strategies that rely solely on gradient information, including the average initial expectation function value remains beneficial. It helps the algorithm select mixer operators that have a stronger influence on the expectation function at the initial stage, thereby accelerating convergence within each optimization round.}
	
	\medskip
	\textcolor{black}{In summary, the numerical results demonstrate that incorporating the average initial expectation function value into the evaluation function of AMA-QAOA+ can improve solution stability in some cases and consistently accelerate the convergence of the expectation function. This leads to reduced quantum resource consumption per run without sacrificing solution quality.
	}

\end{appendix}

\end{document}